\documentclass[aps,reprint]{revtex4-1}

\usepackage{graphicx}% Include figure files
\usepackage{dcolumn}% Align table columns on decimal point
\usepackage{bm}% bold math
\usepackage{epsfig}
\usepackage{amsmath}
\usepackage{amssymb}
\usepackage{xspace}
\begin{document}

%\preprint{APS/123-QED}

\title{The phase behaviour and structure of a fluid confined between competing (solvophobic and solvophilic) walls}

\author{M C Stewart}
\author{R Evans}%
\email{Bob.Evans@Bristol.ac.uk}
\affiliation{H. H. Wills Physics Laboratory, University of Bristol, Bristol BS8 1TL, United Kingdom}%

\date{\today}% It is always \today, today,

\begin{abstract}
We consider a model fluid with long--ranged, $r^{-6}$ (dispersion) interparticle potentials confined between competing parallel walls. One wall is solvophilic and would be completely wet at bulk liquid-gas coexistence $\mu_{co}^-$ while the other is solvophobic and would be completely dry at $\mu=\mu_{co}^+$.  When the wall separation $L$ is large and the system is below the bulk critical temperature $T_C$ and close to bulk liquid--gas coexistence,  a `delocalized interface' or `soft mode' phase forms with a liquid--gas interface near to the centre of the slit; this interacts with the walls via the power-law tails of the interparticle potentials. We use a coarse--grained effective Hamiltonian approach to derive explicit scaling expressions for the Gibbs adsorption, $\Gamma$, the surface tension, $\gamma$, the solvation force, $f_s$ and the total susceptibility $\chi$. These quantities depend on the dimensionless scaling variable $(L/\sigma)^3\beta\delta\mu$, where $\beta=(k_BT)^{-1}$, $\sigma$ is the diameter of the fluid particles and $\delta\mu=\mu-\mu_{co}$ is the chemical potential deviation from bulk coexistence. Using a non-local density functional theory (DFT) we calculate density profiles for the asymmetrically confined fluid at different chemical potentials and, for sufficiently large $L$, confirm the scaling predictions for the four thermodynamic quantities. Since the upper critical dimension for complete wetting with power-law potentials is $<3$ we argue that our (mean-field) scaling predictions should remain valid in treatments that incorporate the effects of interfacial fluctuations. As the wall separation $L$ is decreased at $\mu_{co}$ we predict a capillary evaporation transition from the `delocalized interface' phase to a dilute gas state with just a thin adsorbed film of liquid-like density next to the solvophilic wall. This transition is connected closely to the first order pre--wetting transition which occurs at the solvophilic wall in the semi--infinite system. We compare the phase diagram for the competing walls system with the phase diagrams for the fluid confined between identical solvophilic and identical solvophobic walls. Comparisons are also made with earlier studies of asymmetric confinement for systems with short-ranged forces.
\end{abstract}

\maketitle

\section{Introduction}
\label{sec:int}
Understanding the phase behaviour and equilibrium structure of confined fluids is a subject of intrinsic interest to the statistical mechanics communities and is also a key ingredient in elucidating the microscopic dynamics and flow properties of fluids in microfluidic or nanofluidic devices. When a simple (atomic or molecular) fluid is confined in narrow pores, or channels or capillaries, where the size of the confining dimension is of the order of nanometers, the effects of both finite pore size and substrate fluid interactions (adsorption) can have a profound influence on phase behaviour giving rise to phenomena that have no direct counterpart in bulk \cite{Evans90,GelbGub}. For a real fluid confined by a real structured substrate or in a real porous solid details of the substrate (wall) potentials matter. The nature of the wall--fluid potentials, the roughness of the walls, the atomic corrugations and the geometry of pores or grooves all play a role in determining the microscopic structure of the confined fluid. However, when it comes to phase behaviour one might hope that certain generic features would be captured by simple models of confinement. This is the motivation behind the majority of theoretical and computer simulation efforts to ascertain the nature of phase transitions of fluids confined in idealized geometries; these are often planar (slit--like) or cylindrical model pores with structureless walls representing the substrate \cite{Evans90,GelbGub}. Indeed much of the progress in the field has resulted from detailed studies of even simpler models namely a lattice gas or a nearest-neighbour Ising film of finite thickness $D$ where the spins at opposite walls are subject to (local) surface fields $H_1$ and $H_D$ in addition to the external, bulk magnetic field $H$. Phenomena such as capillary condensation and evaporation, prewetting and the interface delocalization transition have been investigated and elucidated in Ising model studies \cite{BinLanMul} following on from the work of Fisher and Nakanishi in the early eighties  \cite{FishNak,NakFish}. In recent years some focus has shifted to soft matter systems, e.g., colloids and polymers, where the mesoscopic length scales of the `particles' are such that the details of the atomistic structure and roughness of the confining walls should not be relevant. Moreover the large size of the colloidal particles means that these can often be tracked in real space and in time using confocal and video microscopy thereby revealing information about structure. Of course, theoretical techniques developed for atomic fluids are easily adaptable to colloidal systems---the latter having the advantage that by applying suitable chemistry the effective fluid--fluid and wall--fluid interactions can be tailored. For an informative overview of models of soft matter, in particular colloid--polymer mixtures, confined in thin film geometry, i.e., between two parallel walls, see Ref.\ \cite{BinHorVinVir}. The phase behaviour of symmetrical polymer blends in the same confining geometry is reviewed in Ref.\ \cite{MulBin}.

In the present paper we consider confinement of a simple fluid between `competing' planar walls one of which is solvophobic---it is completely wet by the gas phase---whereas the other wall is solvophilic and is completely wet by the liquid phase. The phase behaviour in such a system is very different from that when the two walls are identical \cite{ParryEvPRL,ParryEv}.

Parry and Evans \cite{ParryEvPRL,ParryEv} drew attention to the novel phase equilibria that may arise when a fluid (or Ising magnet) is confined between two perfectly antisymmetric walls. For temperatures below the bulk critical temperature ($T<T_C$) they identified a single `soft mode' phase characterised by a liquid-gas ($+-$ spins in the Ising case) interface which lies parallel to and at the mid-point between the walls when the fluid in the reservoir is at bulk liquid-gas coexistence. In the Landau theory studied by Parry and Evans \cite{ParryEvPRL,ParryEv} the forces are short-ranged and the liquid-gas interface fluctuates freely about the centre of the slit with an exponentially diverging transverse correlation length $\xi_{||}\sim e^{\kappa L/4}$ and the total susceptibility diverges as $e^{\kappa L/2}$, where $L$ is the wall separation. In the original, mean-field Landau analysis \cite{ParryEv} the length scale $\kappa^{-1}$ is the bulk correlation length: $\kappa^{-1}=\xi_b$. Subsequently Parry and co-workers \cite{ParryRev,ParryBoult,BoultParry} used a generalized effective Hamiltonian approach to argue that $\kappa^{-1}=\xi_b(1+\omega/2)$ where the wetting parameter $\omega$ is defined as $\omega=k_BT/(4\pi\xi_b^2\gamma_{lg})$, $\gamma_{lg}$ being the interfacial tension for the free liquid-gas interface or the interfacial stiffness for the Ising case and $\xi_b$ is the true bulk correlation length. This `soft mode' phase persists until a critical temperature $T_{cL}$ where $T_{cL}<T_{W}$, the critical wetting transition temperature at a single wall (for perfectly antisymmetric walls the single wall drying transition temperature $T_D$ is equal to $T_W$). Below $T_{cL}$ symmetry breaking occurs and there is two phase coexistence between states where the liquid-gas interface is bound to either one of the walls. The transition at $T_{cL}$ is sometimes termed an interface localisation-delocalisation transition and can be first or second-order depending on the single wall wetting and drying transitions.

This unusual behaviour for a system subjected to antisymmetric surface fields has been studied in several papers by Binder {\it{et. al.}} \cite{BinLanFerPRL,BinLanFerPRE,BinEvLanFer} using Monte Carlo simulations of a nearest--neighbour lattice gas model (equivalent to an Ising model in magnetic terminology). On decreasing the temperature $T$ below the bulk critical temperature $T_C$ the layer magnetisation $m_n$ (or density $\rho_n=[1-m_n)/2]$) profiles (see, e.g., Fig. 1 in Ref. \cite{BinLanFerPRE}) show the gradual formation of a $+-$ interface between the two bulk phases. The presence of this interface is also evident in the layer susceptibility profile $\chi_{n}=\partial m_n / \partial H$, where $H$ is the external magnetic field equivalent to the reservoir chemical potential deviation $\delta\mu\equiv\mu-\mu_{co}$ in the lattice gas model. $\chi_n$ develops a pronounced peak in the centre of the film. Note that in the Ising simulations the wall separation is denoted $D$ rather than $L$, and the surface fields, acting only on layers $1$ and $D$, satisfy $H_D=-H_1$. For large $L$ the susceptibility is large at $H=0$ because the interface is almost freely floating---the free energy cost for shifting the whole interface is exponentially small. The simulations of Binder {\it{et. al.}} confirmed the Parry and Evans prediction \cite{ParryEv} of a susceptibility which diverges exponentially with $L$. Binder {\it{et. al.}}  \cite{BinLanFerPRL,BinLanFerPRE,BinEvLanFer} also studied the phase transition at $T_{cL}$, where the symmetry is broken and the interface becomes bound to one of the walls rather than delocalised in the centre of the slit. For critical wetting, in agreement with Parry and Evans \cite{ParryEvPRL,ParryEv}, they found that $T_{cL}$ lies close to but below the single wall wetting transition temperature $T_W$. As expected this interface localization--delocalization transition has two--dimensional Ising character. Monte Carlo simulation results for the Ising model are reviewed in \cite{BinLanMul} where they are compared to mean field predictions.

Following these developments in the early and mid nineties the phase behaviour of other types of fluid confined between competitive walls was investigated. M\"uller and co--workers studied binary (symmetric) polymer blends subject to asymmetric surface fields using Monte Carlo and self-consistent field approaches---see e.g., Ref. \cite{MulBinAlb}. A summary of such studies is given in Ref. \cite{MulBin}. It is well-known that sterically stabilized colloidal suspensions with added non--adsorbing polymer exhibit fluid--fluid phase separation into phases that are rich or poor in colloid. The bulk properties of such mixtures are often described by the Asakura--Oosawa--Vrij (AO) model. De Virgiliis and co--workers have investigated the phase behaviour of the AO model fluid between asymmetric confining walls where one wall is completely wet by the colloid rich phase and the other by the polymer rich phase \cite{VirVinHorBin,VirVinHorBinPRE,BinHorVinVir}.

For completeness we also mention recent studies of the (confined) Ising strip in two dimensions subject to various surface fields \cite{DrzMacBarDiet,NowNap,NowNapJPA,VirAlbMulBin,AbrMac}, including the antisymmetric $H_D=-H_1$ case. These reveal `pseudo' transitions in the reduced dimensionality that are analogous to the real transitions occurring in three dimensions. There are also recent Monte Carlo studies of Ising films subject to general asymmetric (local) surface fields \cite{AlbBin}. Note that in Ref.\ \cite{VirAlbMulBin} simulations were performed for the nearest--neighbour Ising strip subject to long--ranged $n^{-3}$ antisymmetric surface fields while Ref.\ \cite{DrzMacBarDiet} performed density matrix renormalization group calculations for the same strip subject to various power--law antisymmetric surface fields. Ref.\ \cite{VirAlbMulBin} presents results for the `soft mode' phase, discussing in detail the effects of capillary wave fluctuations. These effects are much more pronounced in spatial dimension $d=2$ than in $d=3$.

Here we study the behaviour of a simple fluid subject to competing walls at a single temperature $T$ in the `delocalised interface' or `soft mode' phase, i.e., $T_{cL}<T<T_C$. In contrast to previous studies of fluids \cite{BinHorVinVir,VirVinHorBin,VirVinHorBinPRE}, where the interaction potentials were short-ranged, we include the full tail, $-r^{-6}$, of the Lennard-Jones 12-6 pair potential in the fluid-fluid and wall-fluid interactions. As a result the centrally located liquid-gas interface interacts with the two walls via the power-law tails of the interparticle potentials. The overall phase behaviour is expected to be similar to that described above for short-ranged forces but with quantities such as the susceptibility $\chi$ and transverse correlation length $\xi_{||}$ diverging algebraically rather than exponentially \cite{ParryEv}.

In Section \ref{sec:EIP} we employ an effective potential approach in which the excess grand potential is written as a function of the thickness of the drying film $l$ (this is equivalent to the `slab approximation' of Parry and Evans, see Section 2.3 in Ref.\ \cite{ParryEv}). Our findings apply to the general case of competing walls, where one is drying and the other wetting but the walls are not necessarily perfectly antisymmetric. We consider state points both on and away from bulk coexistence $\mu_{co}$. In Section \ref{sec:wetdry} expressions are derived for various thermodynamic quantities including the Gibbs adsorption, $\Gamma$, surface tension, $\gamma$, solvation force, $f_s$ and the total susceptibility $\chi$. For large $L$ these are given by scaling functions of the dimensionless product $(L/\sigma )^3 \beta \delta \mu$ where $\sigma$ is the hard sphere diameter of the fluid particles, $\delta\mu$ is the chemical potential deviation from bulk liquid-gas coexistence and $\beta=(k_BT)^{-1}$. At bulk coexistence, $\delta\mu=0$, the total susceptibility is predicted to diverge as $\chi\propto L^4$. The solvation force, i.e., the excess pressure due to confining the fluid, is predicted to be repulsive: $f_s$ decreases with wall separation as $L^{-3}$, for large $L$.

Capillary condensation (or evaporation) is a well known phenomenon for a fluid confined between two identical parallel walls. In Section \ref{sec:cap} we display phase diagrams (at constant temperature $T=0.8T_C$) for different systems obtained using the effective potential approach. First we consider two identical parallel solvophobic walls and determine the line of evaporation transitions as a function of $\delta\mu$ and inverse wall separation $L^{-1}$. In the second system the walls are both solvophilic. A single wall ($L=\infty$) undergoes a first order wetting transition at $T_W<0.8T_C$ and the accompanying prewetting transition can also occur in the confined fluid where it competes with capillary condensation. We show that both capillary condensation and evaporation are possible in a competing walls system. For the particular system which we study in Sec.\ \ref{sec:DFTres} using density functional theory (DFT) the effective potential approach predicts a capillary evaporation transition which occurs for increasing $\delta\mu$ as the wall separation $L$ is decreased. The line of evaporation transitions crosses bulk coexistence $\mu=\mu_{co}$ at $L=L^{co}_{evap}$.

In Section \ref{sec:sr} we describe sum rules that relate the solvation force to the contact densities at the walls. These are used in checking the accuracy of the DFT calculations.

Section \ref{sec:DFTres} describes numerical results obtained using a non-local DFT. The wall-fluid potentials have been chosen so that at $T=0.8T_C$ the two walls are `antisymmetric', i.e., the Hamaker constant [defined in Eq.\ (\ref{eq:b_app})], for wetting at the solvophilic wall [$b_2$ in Eq.\ (\ref{eq:w2})] is equal to that for drying at the solvophobic wall [$b_1$ in Eq.\ (\ref{eq:w1})]. Results are given for four different wall separations $L/\sigma = 50$, $100$, $250$ and $500$ at temperature $T=0.8T_C$. We find good agreement with the effective potential predictions of Section \ref{sec:EIP}, i.e., scaling is obeyed for all the thermodynamic quantities we consider, provided $L$ is sufficiently large. Density profiles of the two coexisting states at $L_{evap}^{co}=11.4\sigma$, obtained from DFT, are also presented. At $L=100\sigma$ capillary evaporation occurs on the gas side of bulk coexistence at $\beta\delta\mu=-4.13\times10^{-3}$ and we display the coexisting density profiles at this transition. These are almost identical to the profiles at the prewetting transition.

Our results are discussed in Section \ref{sec:distw}, where we also speculate on the phase diagram as temperature is varied. We relate our findings to simulation and theoretical studies of other model fluids and comment on their relevance for experimental results for confined water.

\section{Effective Interfacial Potential Description and its Predictions}
\label{sec:EIP}
Following from the pioneering work of Frumkin and Derjaguin in the 1930's (insightfully reviewed in recent articles by Henderson \cite{Hend05,Hend11}), theories of wetting and condensation phenomena often express the excess grand potential of a confined fluid as a function of a single variational parameter, $l$, the thickness of a wetting film. Here we extend and quantify this effective interfacial potential approach, introduced in Ref.\ \cite{ParryEv} for asymmetric confinement, specializing to the case of power-law attractive potentials.
\subsection{The model}
\label{sec:MOD}
\begin{figure}
\centering
\epsfig{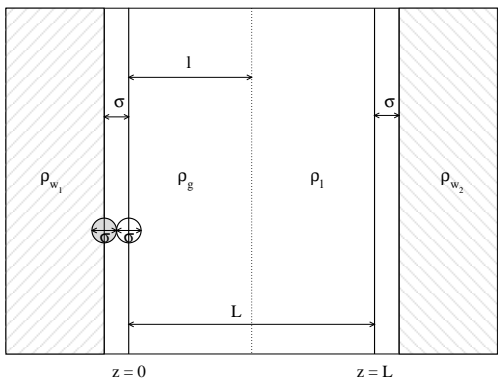}
\caption[The sharp-kink approximation for the density profile of a fluid confined in an asymmetric slit]{Diagram to show the sharp-kink approximation for the density profile $\rho(z)$ of the fluid in an asymmetric slit. The walls are comprised of uniform-density blocks of particles, separated by a distance $L$. The wall-particles interact with the fluid particles with the same potential as the fluid-fluid interactions. A fluid particle is shown at the plane of contact of the fluid with wall$_1$, i.e., at $z=0$. A wall particle (shaded) is shown at the furthest extent of wall$_1$ towards the fluid. There are excluded volumes, of width $dw=\sigma$, in which the density is zero between the fluid and the regions of constant wall density $\rho_{w_1}$ and $\rho_{w_2}$. In the sharp-kink approximation there is a film of gas, width $l$, with the coexisting bulk gas density $\rho_g$ next to (drying) wall$_1$; $\rho_{w_1}<\rho_g$. The remainder of the slit is filled with a film of liquid, width $L-l$, with the coexisting liquid density $\rho_l$. Wall$_2$ is wetting; $\rho_{w_2}>\rho_l$. The density profile is discontinuous at the gas--liquid interface at $z=l$.}
\label{fig:walls}
\end{figure}

The specific model fluid we consider here has a pair potential consisting of a hard core of diameter $\sigma$ plus an attractive tail, taken to be the attractive part of the full ($12,6$) Lennard-Jones (LJ) potential:
\begin{equation}
\phi_{att}(r) =
\begin{cases}
\displaystyle 4\epsilon\left[\left(\frac{\sigma}{r}\right)^{12}-\left(\frac{\sigma}{r}\right)^6\right]\quad &r>r_{min} \\
-\epsilon &r<r_{min}
\end{cases}
\label{eq:Lennard_Jones_Potential}
\end{equation}
where $r_{min}=2^{\frac{1}{6}}\sigma$ and $\epsilon$ is the LJ well-depth. We shall describe this fluid in the context of a density functional theory (DFT)---see Section \ref{sec:DFTres}---that treats the hardcore repulsion within fundamental measures theory and the attractive interactions (\ref{eq:Lennard_Jones_Potential}) in mean--field fashion. In this DFT approach the bulk critical temperature is $k_B T_C / \epsilon=1.415$. The fluid is confined between parallel walls separated by a distance $L$, see Fig.\ \ref{fig:walls}. The wall-fluid potentials are full Lennard-Jones ($9,3$), obtained by integrating the individual Lennard-Jones ($12,6$) wall particle-fluid particle interactions with the same form as (\ref{eq:Lennard_Jones_Potential}) over the volume of the wall, see Eqs.\ (\ref{eq:Lennard_Jones_Potential_Wall})-(\ref{eq:wall_potential_app}).  There is hard repulsion at the contact surface of the wall with the fluid. The strength of the potential exerted by a wall is varied by changing the density of wall particles $\rho_w$. The wall potential is made repulsive by choosing a negative value for $\rho_w$. For simplicity the range of the wall-fluid interparticle potential has been chosen to be the same as that of the fluid-fluid potential, i.e., $\sigma_{wf}=\sigma$. There is an excluded volume of width $dw$ (in this case $dw=\sigma$), measured between the centres of the particles in the surface of the wall and the centres of the fluid particles at the contact surface of the fluid density profile \cite{DietR}, as illustrated in Fig.\ \ref{fig:walls}. The total planar external potential on a fluid particle confined between the two parallel walls is $V(z)=V_{w_1}(z)+V_{w_2}(L-z)$. $V_{w_1}(z)$ is the potential exerted on a fluid particle at $z$ by a single wall$_1$, with the wall$_1$-fluid contact surface at $z=0$. $V_{w_2}(L-z)$ is the potential due to the second wall (wall$_2$) which has its contact surface with the fluid at $z=L$. The wall density $\rho_{w_1}$ can be chosen so that in the semi-infinite system consisting of wall$_1$ in contact with the liquid the wall is completely dry (wet by the gas) at liquid--gas coexistence and conversely the density of wall$_2$, $\rho_{w_2}$, can be chosen to ensure complete wetting by liquid. The fluid between the two walls is in contact with a reservoir with a chemical potential $\mu$ and temperature $T$, i.e., the system is in the grand canonical ensemble.

\subsection{One wall wet and the other dry}
\label{sec:wetdry}
We focus first on the asymmetric confinement where wall$_1$ is completely dry and wall$_2$ is completely wet in the limit $L\rightarrow\infty$. In the sharp-kink approximation the fluid density is constant on both sides of a step function situated at a distance $l$ from wall$_1$, as shown in Fig.\ \ref{fig:walls}. Using this approximation we can write down the excess grand potential per unit area $A$ for the system at bulk liquid-gas coexistence $\mu=\mu_{co}$:
\begin{eqnarray}
\frac{\Omega_{ex}(l;L,\mu_{co})}{A}&\equiv&\frac{\Omega+pV}{A}\nonumber
\\ &=&\gamma_{w_1g}(\mu_{co})+\gamma_{w_2l}(\mu_{co})+\gamma_{lg}\nonumber
\\ &&+w_{w_1g/gl}(l) +w_{w_2l/lg}(L-l) \nonumber
\\ &&+w_{w_1g/w_2l}(L),
\label{eq:GP}
\end{eqnarray}
where $p=p(\mu,T)$ is the pressure in the reservoir, $V$ is the accessible volume and $l$ is the thickness of the gas film. The first three terms correspond to the individual surface tensions of the three interfaces: wall$_1$-gas, wall$_2$-liquid and gas-liquid. The next terms give the interaction potentials between pairs of interfaces. Using the sharp-kink approximation (see Appendix \ref{sec:intint}) we find:
\begin{eqnarray}
w_{w_1g/gl}(l)&=&\frac{b_1}{l^2}+\frac{c_1}{l^3}
\label{eq:w1}
\\
w_{w_2l/lg}(L-l)&=&\frac{b_2}{(L-l)^2}+\frac{c_2}{(L-l)^3},
\label{eq:w2}
\\
w_{w_1g/w_2l}(L)&=&\frac{b_3}{L^2}+\frac{c_3}{L^3}, 
\label{eq:w3}
\end{eqnarray}
where the temperature dependent (Hamaker) coefficients are
\begin{eqnarray}
b_1 &=&(\rho_{g}-\rho_{w_1})(\rho_l-\rho_g)\pi\epsilon\sigma^6/3,
\label{eq:b1}
\\b_2&=&(\rho_l-\rho_{w_2})(\rho_g-\rho_l)\pi\epsilon\sigma^6/3,
\label{eq:b2}
\\
b_3&=&(\rho_{w_1}\rho_l+\rho_{w_2}\rho_g)\pi\epsilon\sigma^6/3,
\label{eq:b3}
\\
c_1 &=& 2\sigma\rho_{w_1}(\rho_l-\rho_g)\pi\epsilon\sigma^6/3,
\label{eq:c1}
\\c_2&=&2\sigma\rho_{w_2}(\rho_g-\rho_l)\pi\epsilon\sigma^6/3,
\label{eq:c2}
\\
c_3&=&-2\sigma(\rho_{w_1}\rho_l+\rho_{w_2}\rho_g)\pi\epsilon\sigma^6/3.
\label{eq:c3}
\end{eqnarray}
$\rho_g$ and $\rho_l$ are the densities of the coexisting gas and liquid at temperature $T$. For simplicity we have set the wall--fluid parameters $\epsilon_{w_1f}=\epsilon_{w_2f}=\epsilon$ in the above equations but since it is the products $\rho_{w_1}\epsilon_{w_1f}$ and $\rho_{w_2}\epsilon_{w_2f}$ which determine the strength of the wall potentials the expressions are still completely general. The lowest order coefficients, $b_1$, $b_2$ and $b_3$ are expected to be correct beyond the sharp-kink approximation but we anticipate that $c_1$, $c_2$ and $c_3$ will be affected by the microscopic details of the density profile.  The direct interaction between the two walls ({\em not} due to the presence of the fluid) is {\em not} included in the excess grand potential (\ref{eq:GP}). As already implied, in the analysis which follows the temperature and wall densities are chosen such that $b_1$ and $b_2$ are both positive, i.e., in the semi-infinite systems consisting of a wall in contact with the liquid at bulk coexistence, wall$_1$ would be completely dry and conversely wall$_2$ would be completely wet.

If the system is not precisely at bulk liquid-gas coexistence then there is an additional term in the excess grand potential because one of the (bulk) phases present is metastable:
\begin{equation}
\Omega_{ex}(l;L,\mu)=\Omega_{ex}(l;L,\mu_{co})+(p-p_m)V_m
\label{eq:GPmu}
\end{equation}
where $p_m$ and $V_m$ are the pressure and volume of the metastable phase at the same chemical potential. On the liquid side of coexistence, i.e, $\mu-\mu_{co}\equiv\delta\mu>0$, and for small $\delta\mu$ the pressure difference is $(p-p_m)=\delta\mu (\rho_l-\rho_g)$ and the volume of the metastable gas is $V_m=lA$. If $\delta\mu$ is negative then it is the liquid phase which is metastable and $(p-p_m)V_m=|\delta\mu|(\rho_l-\rho_g)(L-l)A$. 
\subsubsection{Film thickness and adsorption}
Substituting Eqs.\ (\ref{eq:GP})-(\ref{eq:b3}) into Eq.\ (\ref{eq:GPmu}) we obtain the excess grand potential for the system on the liquid side of coexistence,
\begin{eqnarray}
\frac{\Omega_{ex}(l;L,\mu)}{A}&=&\gamma_{w_1g}(\mu_{co})+\gamma_{w_2l}(\mu_{co})+\gamma_{lg}\nonumber
\\ &\mbox{} &+\frac{b_1}{l^2}+\frac{b_2}{(L-l)^2}+\frac{b_3}{L^2}+\delta\mu(\rho_l-\rho_g)l
\nonumber
\\ &\mbox{} &+\frac{c_1}{l^3}+\frac{c_2}{(L-l)^3}+\frac{c_3}{L^3}\nonumber
\\  &\mbox{} &+O\left[l^{-3},(L-l)^{-3},L^{-3}\right].
\label{eq:GP1}
\end{eqnarray}
The smoothness of the actual profile is responsible for the leading order corrections, $l^{-3}$ etc., to this effective potential; recall $c_1$, $c_2$ and $c_3$ are the sharp-kink values for the coefficients. The corresponding equation for the gas side of coexistence is obtained by replacing the term $\delta\mu(\rho_l-\rho_g)l$ with $\delta\mu(\rho_l-\rho_g)(l-L)$. The equilibrium gas film thickness $l_{eq}$, is found by minimising Eq.\ (\ref{eq:GP1}) w.r.t. $l$,
\begin{equation}
\mbox{}-\frac{2b_1}{l_{eq}^3}+\frac{2b_2}{(L-l_{eq})^3}-\frac{3c_1}{l_{eq}^4}+\frac{3c_2}{(L-l_{eq})^4}+\delta\mu(\rho_l-\rho_g)=0.
\end{equation}
Higher order terms are not displayed.

In the limit $L\rightarrow\infty$, $l_{eq}\rightarrow\infty$ we can neglect terms of order $l_{eq}^{-4}$ and higher. Then the equilibrium gas film thickness can be expressed in terms of a dimensionless scaling function $\Lambda$:
\begin{equation}
\frac{l_{eq}(L,\mu)}{L}=\Lambda\left[\beta\delta\mu (L/\sigma)^3\right],
\label{eq:TWleq}
\end{equation}
where $\Lambda$ satisfies
\begin{equation}
\frac{2b_1}{\Lambda^3}-\frac{2b_2}{(1-\Lambda)^3}=\delta\mu L^3(\rho_l-\rho_g), \quad \delta\mu>0.
\label{eq:TWAdsScale}
\end{equation}
With $\delta\mu$ replaced by $\lvert{\delta\mu}\rvert$ the same scaling function applies on the gas side of coexistence, i.e., for $\delta\mu<0$. Note that in the perfectly antisymmetric case, where $l_{eq}=L/2$ for $\delta\mu=0$, $\Lambda[0]=1/2$ which requires $b_1=b_2$ at the temperature of interest.

We define the (dimensionless) Gibbs excess adsorption, $\Gamma$, in terms of the equilibrium density profile $\rho(z)$:
\begin{equation}
\frac{\Gamma(L,\mu)}{A}\equiv\int_{0^+}^{L^-} {{\rm d}z[\rho(z)-\rho]},
\label{eq:Ads}
\end{equation}
where $\rho\equiv\rho(\mu/T)$ is the reservoir density. On the liquid side of bulk coexistence $\Gamma$ is negative and the gas film thickness in the sharp-kink approximation is given by
\begin{eqnarray}
l_{eq}=\frac{-\Gamma}{A(\rho_l-\rho_g)}, \hspace{10mm} \delta\mu>0.
\label{eq:AdsLiq}
\end{eqnarray}
On the gas side of bulk coexistence $\Gamma$ is positive and the excess adsorption is proportional to the thickness of the liquid film, $L-l_{eq}$,
\begin{eqnarray}
L-l_{eq}=\frac{\Gamma}{A(\rho_l-\rho_g)}, \hspace{10mm}  \delta\mu<0.
\label{eq:AdsGas}
\end{eqnarray}
Although the excess adsorption as defined in Eq.\ (\ref{eq:Ads}) is discontinuous at $\delta\mu=0$ this is because of the jump in the reservoir density. In the sharp-kink approximation we set $\rho=\rho_g$ for $\delta\mu<0$ and $\rho=\rho_l$ for $\delta\mu>0$. The profile and the thickness of the drying film $l_{eq}$ are continuous at $\delta\mu=0$. 

That the adsorption, and other thermodynamic quantities to be described below, should be functions of the scaling variable $\beta\delta \mu (L/\sigma)^3$ is consistent with the heuristic scaling ansatz introduced by Parry and Evans, see Sec.\ 4 in \cite{ParryEv}, in the particular case of dispersion forces.

\subsubsection{Surface Tension}
\label{sec:ST}
Substituting the equilibrium gas film thickness into the excess grand potential (\ref{eq:GP1}), we find that the equilibrium surface excess free energy per unit area, i.e., the surface tension, depends on a scaling function, $\Sigma$:
\begin{eqnarray}
\gamma(L,\mu)&\equiv&\frac{\Omega_{ex}(l_{eq};L,\mu)}{A}
\\
&=&\gamma_{w_1g}(\mu_{co})+\gamma_{w_2l}(\mu_{co})+\gamma_{lg} \nonumber
\\
&&\mbox{} +\frac{1}{L^2}\Sigma\left[\beta\delta \mu (L/\sigma)^3\right]
\label{eq:STscale}
\end{eqnarray}
where
\begin{eqnarray}
\Sigma\left[\beta\delta \mu (L/\sigma)^3\right]&=& \frac{b_1}{\Lambda^2}+\frac{b_2}{(1-\Lambda)^2}+b_3 \nonumber
\\ && +\delta\mu(\rho_l-\rho_g)\Lambda L^3,
\label{eq:STscalefliq}
\end{eqnarray}
when $\delta\mu>0 $ and
\begin{eqnarray}
\Sigma\left[\beta\delta \mu (L/\sigma)^3\right]&=& \frac{b_1}{\Lambda^2}+\frac{b_2}{(1-\Lambda)^2}+b_3 \nonumber
\\ &&+\lvert\delta\mu\rvert(\rho_l-\rho_g)(1-\Lambda) L^3,
\label{eq:STscalefgas}
\end{eqnarray}
when $ \delta\mu<0$. The scaling function $\Sigma$ has dimensions of energy. It is straight forward to show that (\ref{eq:STscale}) and (\ref{eq:AdsLiq}) satisfy the Gibbs adsorption sum rule
\begin{equation}
\frac{\Gamma}{A}=-\left(\frac{\partial\gamma\left(L,\mu\right)}{\partial\mu}\right)_T.
\end{equation}
\subsubsection{Susceptibility}
The total susceptibility $\chi(L,\mu)$ measures the response of the confined fluid to changes in the chemical potential $\mu$. It is defined as
\begin{equation}
\chi(L,\mu)\equiv-\frac{1}{A}\left(\frac{\partial\Gamma}{\partial\mu}\right)_{L,T}.
\end{equation}
Within the sharp-kink approximation
\begin{equation}
\chi(L,\mu)=(\rho_l-\rho_g)\left(\frac{\partial l_{eq}}{\partial\mu}\right)_{L,T},
\end{equation}
which is continuous at $\delta\mu=0$. From Eqs.\ (\ref{eq:TWleq}, \ref{eq:AdsLiq}, \ref{eq:AdsGas}) we find to leading order
\begin{eqnarray}
\chi(L,\mu)&=&-(\rho_l-\rho_g)^2L^4\left(\frac{6b_1}{\Lambda^4}+\frac{6b_2}{(1-\Lambda)^4}\right)^{-1}
\label{eq:SusScale}
\\
&=& L^4  \; C\left[\beta\delta\mu (L/\sigma)^3\right]
\nonumber
\end{eqnarray}
where $C\left[\beta\delta\mu (L/\sigma)^3\right]$ is a scaling function with dimension length$^{-6}.$energy$^{-1}$.
The susceptibility at bulk coexistence $\delta\mu=0$ is
\begin{equation}
\chi(L,\mu_{co})=\frac{-L^4(\rho_l-\rho_g)^2}{6\left(b_1^{1/3}+b_2^{1/3}\right)^4\left(b_1^{-1/3}+b_2^{-1/3}\right)}.
\label{eq:SusCo}
\end{equation}
In the perfectly antisymmetric situation, $b_1=b_2$,
\begin{equation}
\chi(L,\mu_{co})=\frac{-L^4(\rho_l-\rho_g)^2}{192b_1}.
\label{eq:SusCoAs}
\end{equation}

The local susceptibility measures the change in the fluid density at each point in the density profile $\rho(z)$ as the chemical potential $\mu$ is varied:
\begin{equation}
\chi(z;L,\mu)\equiv-\left(\frac{\partial \rho(z)}{\partial \mu}\right)_T.
\label{eq:LocSusDef}
\end{equation}
The local susceptibility in the region of the liquid-gas interface is expected to be large in magnitude as the liquid-gas interface shifts with $\delta\mu$ whilst maintaining its shape. An increase in the chemical potential $\mu$ causes a decrease in the gas film thickness, $l_{eq}\rightarrow l_{eq}+\delta l_{eq}$, and the density profile near the liquid-gas interface is translated by a (negative) amount $\delta l_{eq}$ in the $z$-direction. The local susceptibility near to the liquid-gas interface is therefore
\begin{equation}
\chi(z;L,\mu)\approx\left(\frac{\partial l_{eq}}{\partial \mu}\right)_{L,T}\left(\frac{\partial \rho(z)}{\partial z}\right), \quad z\sim l_{eq}
\label{eq:LocSus}
\end{equation}
which also diverges as $L^4$. Away from the region near to the liquid-gas interface the density profile is expected to be largely unaffected by small changes in the chemical potential. Correspondingly the local susceptibility is expected to be smaller in these regions than for $z\sim l_{eq}$.
\subsubsection{Solvation force}
\label{sec:fsL}
An important quantity in the theory of confined fluids is the solvation force, $f_S(L,\mu)$, defined as the force per unit area, arising from the presence of the fluid, that must be exerted on the walls to maintain them at separation $L$. $f_s$ does not include any direct interaction between the walls. Although $f_s$ is referred to as a ``force'' it has dimensions of pressure and thermodynamically should be thought of as the excess pressure in the fluid due to confinement, i.e.,
\begin{equation}
f_s(L,\mu)\equiv-\frac{1}{A} \left [\frac{\partial \Omega(l_{eq};L,\mu)}{\partial L} \right ]_{\mu,T,A}-p.
\label{eq:f_sp}
\end{equation}
 It can be calculated from the change in surface tension $\gamma(L,\mu)$ with wall separation $L$,
\begin{equation}
f_s(L,\mu)=-\left[\frac{\partial \gamma(L,\mu)}{\partial L}\right]_{\mu,T,A}.
\label{eq:f_s}
\end{equation}
The solvation force between our asymmetric walls is found from (\ref{eq:GP1})and (\ref{eq:TWleq}); to leading order
\begin{equation}
f_s(L,\mu)=\frac{1}{L^3}F_s\left[\beta\delta\mu (L/\sigma)^3\right]
\label{eq:f_sscale}
\end{equation}
where $F_s$ is another scaling function given by:
\begin{equation}
F_s\left[\beta\delta\mu (L/\sigma)^3\right]=\frac{2b_2}{(1-\Lambda)^3}+2b_3.
\label{eq:FsScale}
\end{equation}
$F_s$ has the dimensions of energy.
\subsection{Capillary condensation, evaporation and prewetting}
\label{sec:cap}
\begin{figure}
\centering
\epsfig{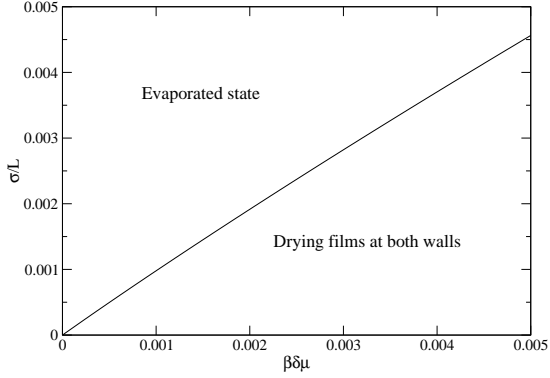}
\caption[Phase diagram for a fluid confined between two identical {\em solvophobic} walls]{Phase diagram for a fluid confined between two identical {\em{solvophobic}} walls at $T=0.8T_C$. The wall-fluid interparticle potentials are given by Eq.\ (\ref{eq:wall_potential_app}) with $\rho_{w1}\epsilon_{w_1f}=-1.108\epsilon\sigma^{-3}$. This choice of parameters ensures that $b_1>0$ and therefore complete drying would occur at an isolated wall$_1$. Eq.\ (\ref{eq:IDevap}) is used to calculate the coexistence line between the two phases. The sharp-kink value was assumed for the coefficient $b_1$ and the liquid-gas surface tension $\gamma_{lg}$ was calculated using DFT.}
\label{fig:EvapId}
\end{figure}

Before we examine possible phase transitions in the fluid confined between asymmetric walls it is helpful to recall the transitions that can occur when the two walls are identical. Firstly we consider the capillary evaporation transition that may take place in a fluid confined between two solvophobic walls. For a fluid with $\delta\mu>0$ confined between two planar drying walls we equate the excess grand potential of the evaporated state filled by gas, $\Omega_{ex}^g(L,\mu)$, with that of the state in which the slit is  filled mainly with liquid apart from a layer of gas, thickness $l_{eq}$ at each wall, $\Omega_{ex}^l(l_{eq};L,\mu)$. Within the sharp-kink approximation the excess grand potentials of these two states are given by:
\begin{equation}
\frac{\Omega_{ex}^g(L,\mu)}{A}=2\gamma_{w_1g}(\mu)+\delta\mu(\rho_l-\rho_g)L+O(L^{-2})
\label{eq:Omega_g_L_mu}
\end{equation}
and
\begin{eqnarray}
\frac{\Omega_{ex}^l(l;L,\mu)}{A}&=&2\gamma_{w_1g}(\mu)+2\gamma_{lg}+\frac{2b_1}{l^2} +2\delta\mu(\rho_l-\rho_g)l\nonumber
\\
&&\mbox{}+O[(L-2l)^{-2}]+O(L^{-2}).
\label{eq:Omega_l_L_mu}
\end{eqnarray}
Again the equilibrium thickness $l_{eq}$ is obtained by minimising w.r.t. $l$. $\gamma_{w_1g}(\mu)$ is the surface tension of the wall-gas interface and $b_1$ is the Hamaker constant for drying at a single wall, given by Eq.\ (\ref{eq:b1}). The corrections of order $L^{-2}$ in Eqs.\ (\ref{eq:Omega_g_L_mu}) and (\ref{eq:Omega_l_L_mu}) refer to the interaction potential between the two wall-gas interfaces; those of order $(L-2l)^{-2}$ in (\ref{eq:Omega_l_L_mu}) arise from the interaction potential between the two gas-liquid interfaces. Capillary evaporation occurs when the two excess grand potentials are equal:
\begin{equation}
(\rho_l-\rho_g)\delta\mu=\frac{2\gamma_{lg}}{\left[L_{evap}-3l_{eq}\left(\delta\mu\right)\right]}
\label{eq:IDevap}
\end{equation}
where $L_{evap}$ is the wall separation at evaporation. The factor of $3$ in the denominator reflects directly the effects of power--law (dispersion) forces---see \cite{EvansMarc}. For short-range fluid-fluid and wall-fluid potentials $(\rho_l-\rho_g)\delta\mu=2\gamma_{lg}/\left[L_{evap}-2l_{eq}(\delta\mu)\right]$. Figure \ref{fig:EvapId} displays the phase diagram, calculated using Eq.\ (\ref{eq:IDevap}), for the fluid between identical, parallel solvophobic walls as the wall separation $L$ and chemical potential deviation from coexistence $\delta\mu$ are varied. As $\delta\mu$ increases the wall separation at evaporation decreases reflecting the increased free energy cost ($\propto \delta\mu$) of a volume of fluid at the gas density.

\begin{figure}
\centering
\epsfig{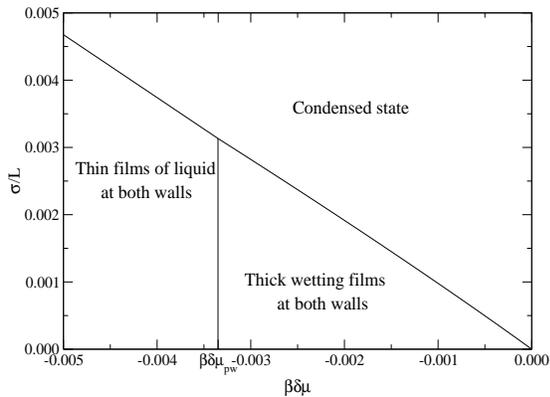}
\caption[Phase diagram for a fluid confined between two identical {\em solvophilic} walls]{Phase diagram for a fluid confined between two identical {\em solvophilic} walls at $T=0.8T_C$. The wall-fluid interparticle potentials are given by Eq.\ (\ref{eq:wall_potential_app}) with $\rho_{w2}\epsilon_{w_2f}=1.723\epsilon\sigma^{-3}$. This choice of parameters ensures that $b_2>0$ and therefore complete wetting would occur at an isolated wall$_2$. The coexistence lines between the different states were calculated using Eqs.\ (\ref{eq:thin}), (\ref{eq:thick}) and (\ref{eq:cond}), assuming the sharp-kink value for the coefficient $b_2$. DFT was used to calculate the surface tensions for the liquid-gas $\gamma_{lg}$, wall$_2$-liquid $\gamma_{w_2l}(\mu_{co})$ and wall$_2$-gas $\gamma^*_{w_2g}(\mu_{co})$ interfaces. Note that the wall$_2$-gas surface tension $\gamma^*_{w_2g}(\mu_{co})$ is the excess grand potential per unit area of the thin film state and refers to a metastable state at bulk coexistence $\mu_{co}$. The coexistence values for the wall$_2$-fluid surface tensions were used in the calculations; we approximated $\gamma_{w_2l}(\mu)\approx\gamma_{w_2l}(\mu_{co})$ and $\gamma^*_{w_2g}(\mu)\approx\gamma^*_{w_2g}(\mu_{co})$. $\mu_{pw}$ denotes the chemical potential at pre-wetting, $L=\infty$. In this approximation the thin--thick transition is independent of $L$.}
\label{fig:Cond}
\end{figure}

More complex phase behaviour is predicted for our fluid confined between two identical solvophilic walls with the same wall-fluid potential as wall$_2$ in the asymmetric system. In order to understand the transitions that occur in the confined fluid it is instructive to first consider the wetting transition that would take place in the semi-infinite system, $L=\infty$, at wall$_2$ on increasing the temperature along the bulk coexistence curve on the gas side $\mu=\mu_{co}^-(T)$. In contrast to the solvophobic wall$_1$ which is dry at all temperatures, at wall$_2$ there is a first-order wetting transition at a temperature $T_W$, i.e., the thickness of the adsorbed liquid film jumps from being finite below $T_W$ to infinite (macroscopic) above $T_W$. Above $T_W$ the first-order thin--thick transition occurs off coexistence at a chemical potential deviation from coexistence $\delta\mu_{pw}(T)\equiv\mu_{pw}(T)-\mu_{co}<0$. In this `prewetting' transition the film thickness jumps from a thin to a thick but finite value. Prewetting persists up to the prewetting critical temperature $T_C^{pw}$. The temperature at which we perform our DFT investigations, $T=0.8T_C$, is above the wetting temperature but below the prewetting critical temperature, i.e., $T_W<0.8T_C<T_C^{pw}$. Consequently we observe a thin--thick film transition in the semi-infinite system. This prewetting transition may also take place in the symmetrically confined fluid, at a very similar value of chemical potential to that of the transition at a single wall when $L$ is large. The excess grand potential of the thin film state is:
\begin{equation}
\frac{\Omega_{ex}^{thin}(L,\mu)}{A}=2\gamma_{w_2g}(\mu)+O(L^{-2})
\label{eq:thin}
\end{equation}
where $\gamma_{w_2g}(\mu)$ is the surface tension of the wall$_{2}$--gas interface (the thin liquid film state) for the semi-infinite fluid at a single wall. The excess grand potential of the thick liquid film state is:
\begin{eqnarray}
\frac{\Omega_{ex}^{thick}(l_{liq};L,\mu)}{A}&&=2\gamma_{w_2l}(\mu)+2\gamma_{lg}+\frac{2b_2}{l_{liq}^2}
\\
\mbox{}+&&2|\delta\mu|(\rho_l-\rho_g)l_{liq}+O[(L-2l_{liq})^{-2}]  \nonumber
\label{eq:thick}
\end{eqnarray}
where $\gamma_{w_2l}(\mu)$ is the surface tension of the wall-liquid interface, $l_{liq}$ is the thickness of the wetting (liquid) film and $b_2$ is the Hamaker constant for wetting at a single wall, given by Eq.\ (\ref{eq:b2}). The thin-thick film transition occurs when $\Omega_{ex}^{thin}(L,\mu)=\Omega_{ex}^{thick}(l_{liq};L,\mu)$. However the thin-thick film transition is now in competition with capillary condensation. The excess grand potential of the condensed liquid state is:
\begin{equation}
\frac{\Omega_{ex}^{cond}(L,\mu)}{A}=2\gamma_{w_2l}(\mu)+|\delta\mu|(\rho_l-\rho_g)L+O(L^{-2}).
\label{eq:cond}
\end{equation}
The equilibrium state at any given wall separation $L$ and chemical potential $\mu$ is the state with the lowest excess grand potential. The phase diagram, calculated using Eqs.\ (\ref{eq:thin}), (\ref{eq:thick}) and (\ref{eq:cond}), is shown in Fig.\ \ref{fig:Cond}. At constant (large) $L$, increasing the chemical potential results in two first-order transitions on the approach to bulk coexistence---first a thin-thick film prewetting transition at $\delta\mu\approx\delta\mu_{pw}$, independent of $L$ in the present approximation, and then a transition from a state with thick wetting films at both walls and gas in the centre to a condensed state in which the slit is completely filled with liquid. At small $L$, i.e., $\sigma/L\gtrsim0.00465$, there is a single transition from a state with thin liquid films at both walls to the condensed state. There is a triple point where the prewetting transition intersects the condensation transition line. The genesis of such a triple point was described in earlier papers based on a simple DFT \cite{EvansMarcPRA} and a lattice gas model \cite{BrunMarEv}.

Now we return to the asymmetrically confined fluid. We have already seen that for $T$ above the wetting and drying temperatures of the two walls, when $L$ is large and the chemical potential is close to bulk coexistence, the slit contains a region of gas phase next to the drying wall and a region of liquid phase next to the wetting wall with a liquid-gas interface somewhere near the centre. We shall refer to this phase as the `delocalised interface state' (i). Let us consider the prewetting transition that occurs at the solvophilic wall in the semi-infinite system. For $\delta\mu<\delta\mu_{pw}$, apart from a thin liquid film at the solvophilic wall, the system contains mainly fluid at the bulk gas density. The situation in the finite $L$ system at $\delta\mu<\delta\mu_{pw}$ is expected to be quite similar---the presence of the solvophobic wall will not alter the fluid density near to the solvophilic wall. The excess grand potential of this phase, which we shall refer to as the evaporated state (g), is given by
\begin{equation}
\frac{\Omega_{ex}^{g}(L,\mu)}{A}=\gamma_{w_1g}(\mu)+\gamma_{w_2g}(\mu)+\frac{b_{g}}{L^2} +O(L^{-3})
\label{eq:GP_evgas}
\end{equation}
when $\delta\mu<0$ and
\begin{eqnarray}
\frac{\Omega_{ex}^{g}(L,\mu)}{A}&=&\gamma_{w_1g}(\mu)+\gamma_{w_2g}(\mu)+\frac{b_{g}}{L^2}+\delta\mu(\rho_l-\rho_g)L  \nonumber
\\&& \mbox{}+O(L^{-3})
\label{eq:GP_evliq}
\end{eqnarray}
when $\delta\mu>0$, where $\gamma_{w_2g}(\mu)$ is the surface tension of the thin film state in the semi-infinite fluid at wall$_2$. (Note that when $\delta\mu>\delta\mu_{pw}$ this wall$_2$-thin film state is metastable). The $b_g/L^2$ terms in Eqs.\ (\ref{eq:GP_evgas}) and (\ref{eq:GP_evliq}) arise from the interaction between the two wall-gas interfaces; the coefficient for this term is given by $b_{g}=(\rho_{w_1}+\rho_{w_2}-\rho_g)\rho_g\pi\epsilon\sigma^6/3$. There is an extra term, $\delta\mu(\rho_l-\rho_g)L$, in the excess grand potential of the evaporated state (\ref{eq:GP_evliq}) for $\delta\mu>0$ because the bulk gas phase is metastable. At the prewetting transition in the semi-infinite system the wetting film thickness increases discontinuously so that the density profile has separate wall$_2$-liquid and liquid-gas interfaces. In the asymmetric system the corresponding state has a thick film of liquid at wall$_2$ separated from the gaseous region next to wall$_1$ by a liquid-gas interface. This is what we term the delocalised interface state $(i)$. The excess grand potential of this state (see Section \ref{sec:ST}) is what we calculated already
\begin{eqnarray}
\frac{\Omega_{ex}^i(l_{eq};L,\mu)}{A}&&=\gamma_{w_1g}(\mu)+\gamma_{w_2l}(\mu)+\gamma_{lg} \nonumber
\\
&&\mbox{}+\frac{1}{L^2}\Sigma\left[\beta\delta\mu(L/\sigma)^3\right]+O(L^{-3}),
\label{eq:GP_mixed}
\end{eqnarray}
where $\Sigma[\beta\delta\mu(L/\sigma)^3]$ is a scaling function, given by Eqs.\ (\ref{eq:STscalefliq}) and (\ref{eq:STscalefgas}). Coexistence between the evaporated state and the delocalised interface state occurs when 
\begin{equation}
\Omega_{ex}^g(L,\mu_{evap})=\Omega_{ex}^i(l_{eq};L,\mu_{evap})
\label{eq:evap}
\end{equation}
 where $\mu_{evap}$ is the chemical potential at evaporation.
\begin{figure}
\centering
\epsfig{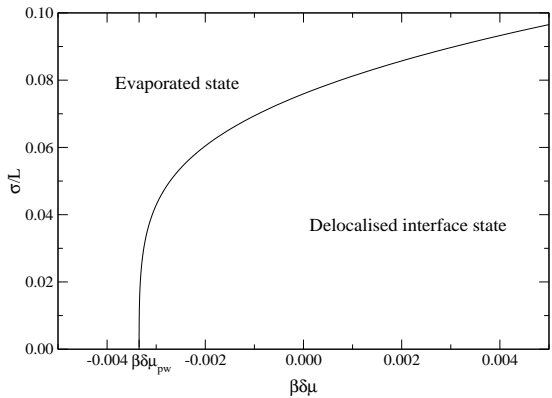}
\caption[Phase diagram for a fluid confined between {\em asymmetric} walls]{Phase diagram for a fluid confined between {\em asymmetric} walls at $T=0.8T_C$. The wall-fluid interparticle potentials are given by Eq.\ (\ref{eq:wall_potential_app}) with $\rho_{w1}\epsilon_{w_1f}=-1.108\epsilon\sigma^{-3}$ and $\rho_{w2}\epsilon_{w_2f}=1.723\epsilon\sigma^{-3}$. These parameters were chosen to ensure complete drying at isolated wall$_1$ and complete wetting at wall$_2$ at a temperature of $T=0.8T_C$. They ensure $b_1=b_2$ at this temperature. Eq.\ (\ref{eq:evap}) was used to calculate the coexistence line between the two phases. The sharp-kink values were assumed for the coefficients $b_1$, $b_2$, $b_3$ and $b_g$. DFT was used to calculate the surface tensions for the liquid-gas $\gamma_{lg}$, wall$_2$-liquid $\gamma_{w_2l}(\mu_{co})$ and wall$_2$-gas $\gamma^*_{w_2g}(\mu_{co})$ interfaces. Note that the wall$_2$-gas surface tension $\gamma^*_{w_2g}(\mu_{co})$ is the excess grand potential per unit area of the thin film state and refers to a metastable state at bulk coexistence $\mu_{co}$. Again, we used the approximations $\gamma_{w_2l}(\mu)\approx\gamma_{w_2l}(\mu_{co})$ and $\gamma^*_{w_2g}(\mu)\approx\gamma^*_{w_2g}(\mu_{co})$. $\mu_{pw}$ denotes the chemical potential for prewetting at a single solvophilic wall$_2$.}
\label{fig:EvapOpp}
\end{figure}
The phase diagram as a function of $\delta\mu$, the chemical potential deviation from bulk coexistence, and inverse wall separation $L^{-1}$ is displayed in Fig. \ref{fig:EvapOpp}. At large $L$ evaporation occurs close to the chemical potential for prewetting at the isolated solvophilic wall, i.e., as $L\rightarrow\infty$, $\delta\mu_{evap}\rightarrow\delta\mu_{pw}$ and in Sec.\ \ref{sec:DFTres} we shall find the coexisting density profiles near to the solvophilic wall at the evaporation transition are very similar to the thin and thick film density profiles for the semi-infinite fluid at the solvophilic wall. One observes in Fig.\ \ref{fig:EvapOpp} that as $L$ is decreased this transition moves to higher chemical potential, i.e., is closer to bulk coexistence, as the interaction of the liquid-gas interface with the solvophobic wall becomes more significant. Evaporation occurs at bulk liquid-gas coexistence ($\delta\mu=0$) for a wall separation of
\begin{equation}
L_{evap}^{co}=\left[\frac{\left(b_1^{1/3}+b_2^{1/3}\right)^3+b_3-b_g}{\gamma_{w_2g}^*(\mu_{co})-\gamma_{w_2l}(\mu_{co})-\gamma_{lg}}\right]^{1/2}.
\label{eq:Levap}
\end{equation}
For $L<L_{evap}^{co}$ the evaporation transition occurs on the liquid side of bulk coexistence---as for the evaporation transition in the slit with identical solvophobic walls (see Fig.\ \ref{fig:EvapId}). However, for a given $L$ evaporation occurs at a much lower value of $\delta\mu$ in the asymmetric case.

In the asymmetric system that we have studied using DFT (see Section \ref{sec:DFTres} below) the solvophilic wall exhibits a first order wetting transition and, therefore, prewetting below the temperature of our investigations ($T=0.8T_C$) whereas the solvophobic wall is purely repulsive and is therefore dry at all temperatures. However, if the solvophobic wall potential was constructed such that there was a first-order drying transition at some temperature $T_D$ then a condensation transition, similar to the evaporation transition described above and connected to the predrying transition in the semi-infinite fluid at a single solvophobic wall, would be possible for temperatures $T_D<T<T_C^{pd}$ (where $T_{C}^{pd}$ is the predrying critical temperature).
\section{Sum rules for the confined fluid}
\label{sec:sr}
Recall that the total external potential acting on the fluid due to the two walls is
\begin{equation}
V(z;L)=V_{w1}(z)+V_{w2}(L-z),
\label{eq:wallpot}
\end{equation}
and consider
\begin{eqnarray}
\frac{1}{A}\left(\frac{\partial\Omega}{\partial L}\right)_{T}&=&\int{{\rm d}\,z \left(\frac{\delta\Omega}{\delta [\mu-V(z;L)]}\right)_{T}\left(\frac{\partial [\mu-V(z;L)]}{\partial L}\right)_{T}} \nonumber
\\
&=&-\int{{\rm d}z \, \rho_L(z)\left(\frac{\partial[\mu-V(z;L)]}{\partial L}\right)_{T}} \nonumber
\\
&=&\int{{\rm d}z \, \rho_L(z)V'_{w2}(L-z)}
\label{eq:sr1}
\end{eqnarray}
where we have used the following result for the functional derivative of the grand potential w.r.t. an external potential $V_{ex}(z)$:
\begin{equation}
\left(\frac{\delta\Omega}{\delta[\mu-V_{ex}(z)]}\right)_T=-\rho(z)
\end{equation}
$\rho_L(z)$ in (\ref{eq:sr1}) is the density profile of the fluid between the two walls at separation $L$. Let $V_{w2}(z)=V_H(z)+V_{att2}(z)$ where the first term is the potential due to a hard wall at $z=0$ and the second term is the attractive Lennard-Jones part of the wall potential. Then one can easily show
\begin{equation}
\frac{1}{A}\left(\frac{\partial\Omega}{\partial L}\right)_{T}=-\beta^{-1}\rho_{w2,L}+\int_{0^+}^{L^-}{{\rm d}z \, \rho_L(z)V'_{att2}(L-z)}
\label{eq:wall2}
\end{equation}
where $\rho_{w_2,L}$ is the contact density of the confined fluid at wall$_2$ for wall separation $L$. Exchanging the two walls does not make any difference to the grand potential so we also have:
\begin{equation}
\frac{1}{A}\left(\frac{\partial\Omega}{\partial L}\right)_{T}=-\beta^{-1}\rho_{w1,L}+\int_{0^+}^{L^-}{{\rm d}z \, \rho_L(z)V'_{att1}(z)}.
\label{eq:wall1}
\end{equation}
At an isolated single hard wall we have the well--known result relating the pressure $p$ of the fluid in the reservoir, far from the walls, to the density profile:
\begin{equation}
p=\beta^{-1}\rho_{w,\infty}-\int_{0^+}^{\infty}{{\rm d}z \, \rho_{\infty}(z)V'_{att}(z)}
\label{eq:singwall}
\end{equation}
where $\rho_{w,\infty}$ is the contact density at the single wall and $\rho_\infty(z)$ is the semi-infinite density profile at that wall.
Substituting (\ref{eq:singwall}) and either (\ref{eq:wall2}) or (\ref{eq:wall1}) into the definition for the solvation force Eq.\ (\ref{eq:f_sp}) we obtain:
\begin{eqnarray}
\beta f_s(L)&=&\rho_{w1,L}-\rho_{w1,\infty} - \beta\left[\int{{\rm d}z \, \rho_L(z)V'_{att1}(z)} \right. \nonumber
\\ && \hspace{26mm}\mbox{} \left. -\int{{\rm d}z \, \rho_\infty(z)V'_{att1}(z)}\right]
\label{eq:Fs1}
\\
&=&\rho_{w2,L}-\rho_{w2,\infty} - \beta\left[\int{{\rm d}z \, \rho_L(z)V'_{att2}(L-z)} \right. \nonumber
\\ && \hspace{28mm} \mbox{} \left. -\int{{\rm d}z \, \rho_\infty(z)V'_{att2}(z)}\right]. \hspace{10mm}
\label{eq:Fs2}
\end{eqnarray}
These results for the solvation force can be regarded as sum rules. In an exact treatment computing $f_s(L)$ from (\ref{eq:Fs1}) or (\ref{eq:Fs2}) must give identical results to those obtained from evaluating the derivative of the excess grand potential, i.e., (\ref{eq:f_sp}) or (\ref{eq:f_s}). Non-local DFT treatments satisfy these sum rules. Thus within the context of DFT, the sum rules provide a stringent test of the accuracy of calculations. Provided we know the density profile of the fluid at one of the single walls then Eq. (\ref{eq:Fs1}) or Eq. (\ref{eq:Fs2}) allows us to calculate the solvation force using the density profile for the confined fluid at just one value of the wall separation $L$ rather than performing the derivative in Eq.\ (\ref{eq:f_sp}) numerically. Comparing results from both Eqs. (\ref{eq:Fs1}) and (\ref{eq:Fs2}) provides a consistency check for our DFT calculations below (Sec.\ \ref{sec:DFTres}). Such checks are important given the small free energy differences and therefore the resulting sensitivity of the phase transitions and scaling functions to the numerics.

\section{DFT Results}
\label{sec:DFTres}
 In this section we present numerical results, obtained using a fully microscopic DFT approach, for a fluid confined between planar opposing walls. The fluid-fluid and wall-fluid interaction potentials are described earlier in Section \ref{sec:MOD}. The excess hard sphere part of the free energy functional ${\cal F}_{ex}^{hs}$ was treated by means of Rosenfeld's fundamental measures theory \cite{Rosen89} and the attractive part of the fluid-fluid interaction potential was treated in mean-field fashion. The functional used was the same as in Refs.\ \cite{StewEv,StewEvJPCM} where it was used to describe drying at the surface of a sphere. The grand potential functional is
\begin{eqnarray}
\Omega_V[\rho]&=&{\cal F}_{id}[\rho]+{\cal F}_{ex}^{hs}[\rho]  \nonumber
\\
&& \mbox{}+\frac{1}{2}\int{\int{{\rm d}{\mathbf r}_1{\rm d}{\mathbf r}_2\ \rho({\mathbf r}_1)\rho({\mathbf r}_2)\phi_{att}(\lvert{\mathbf r}_1-{\mathbf r}_2\rvert)}} \nonumber
\\
&&\mbox{}+\int{\rho({\mathbf r})(V({\mathbf r})-\mu)\,{\rm d}{\mathbf r}}.
\label{eq:FGP}
\end{eqnarray}
where the density profile $\rho({\mathbf r})=\rho(z)$ and the external potential is given by (\ref{eq:wallpot}): $V({\mathbf r})\equiv V(z) = V(z;L)$. ${\cal F}_{id}[\rho]$ is the Helmholtz free energy functional for the ideal gas. $\phi_{att}$ is given in (\ref{eq:Lennard_Jones_Potential}).  The equilibrium density profile was found by minimising the grand potential functional $\Omega_V[\rho]$ and the corresponding equilibrium excess grand potential $\Omega_{ex}$ was calculated. The numerical methods employed to obtain the equilibrium density profile are described in detail in Chapter 3 of Ref.\ \cite{StewThesis}. 

 The bulk coexistence curve calculated from the functional (\ref{eq:FGP}) has the standard mean-field form with the critical temperature $k_BT_c/\epsilon=1.415$ and the critical density $\rho_c\sigma^3=0.2457$. All our results were obtained at the same temperature $T=0.8T_C$. Our choice of parameters, $\rho_{w1}\epsilon_{w_1f}=-1.108\epsilon\sigma^{-3}$ and $\rho_{w2}\epsilon_{w_2f}=1.723\epsilon\sigma^{-3}$, ensures complete drying at an isolated wall$_1$ and complete wetting at wall$_2$ and also results in equal coefficients $b_1=b_2$ in the effective interfacial potential (see Eqs.\ \ref{eq:GP}, \ref{eq:b1} and \ref{eq:b2}) at this temperature. 

\begin{figure}
\centering
\epsfig{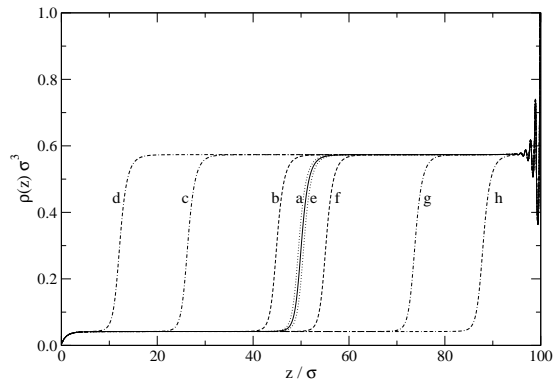}
\caption[Density profiles for a fluid between asymmetric walls, with wall separation of $L=100\sigma$, at various chemical potentials]{Density profiles $\rho(z)$ at various chemical potentials for the asymmetric slit with a wall separation of $L=100\sigma$ and temperature $T=0.8T_C$. The full line shows the profile at bulk liquid-gas coexistence, $\delta\mu=0$. Because $b_1=b_2$ the gas-liquid interface is located at the mid-point between the walls for $\delta\mu=0$. Profiles a-d correspond to chemical potentials on the liquid side of coexistence, $\beta\delta\mu=10^{-6}$, $10^{-5}$, $10^{-4}$ and $10^{-3}$ respectively. Profiles e-h occur on the gas side of coexistence for $\beta\delta\mu=-10^{-6}$, $-10^{-5}$, $-10^{-4}$ and $-10^{-3}$.}
\label{fig:profs100dmu}
\end{figure}

Figure \ref{fig:profs100dmu} shows density profiles of the confined fluid, with wall separation $L=100\sigma$, at liquid-gas coexistence and also for the fluid on both sides of coexistence. The walls are chosen to be `antisymmetric' in that the leading order term in the interaction potentials between each wall-fluid interface and the liquid-gas interface is the same (i.e., $b_1=b_2$). Consequently the liquid-gas interface lies in the centre of the slit at bulk coexistence $\delta\mu=0$. If we increase the reservoir chemical potential away from coexistence into the liquid region of the phase diagram so that $\delta\mu$ is positive then the gas-liquid interface moves away from the centre and towards the drying wall, decreasing the volume of the now metastable gas phase. If $\delta\mu$ is negative so that the reservoir is in the stable gas phase then the volume of liquid phase shrinks and the gas-liquid interface is closer to the wetting wall. The shape of the density profiles near to the walls is largely unaffected by these changes in the chemical potential. Packing effects in the liquid adsorbed at the wetting wall result in oscillations in the density profile in the close vicinity of the wall. The density profile is monotonically decreasing on approaching the drying wall, with a contact density slightly below the bulk gas value \cite{StewEv}.

\begin{figure}
\centering
\epsfig{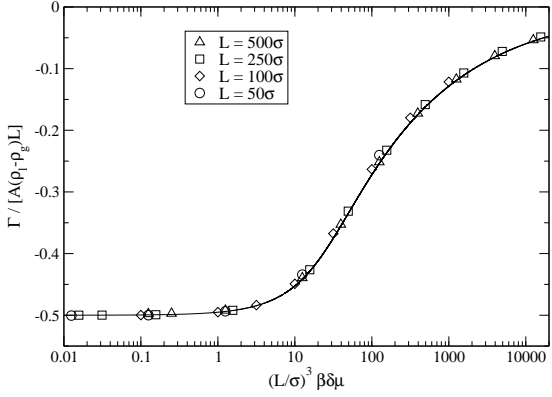}
\caption[Plot to illustrate adsorption scaling in the asymmetric slit on the liquid side of bulk coexistence]{Plot to illustrate adsorption scaling in the asymmetric slit on the liquid side of bulk coexistence, i.e., $\delta\mu>0$, where the adsorption is related to the thickness of the gas film by $l_{eq}=-\Gamma/[A(\rho_l-\rho_g)]$. The full line shows the prediction of the effective potential approach, i.e., the function $-\Lambda\left[\beta\delta\mu(L/\sigma)^3\right]$ valid as $L\rightarrow\infty$, given by Eq.\ (\ref{eq:TWAdsScale}). The symbols are DFT results for various wall separations.}
\label{fig:adslnliq}
\end{figure}

In order to test the effective potential prediction for the film thickness scaling function, $\Lambda \left[\beta\delta\mu (L/\sigma)^3\right]$, Eqs.\ (\ref{eq:TWleq}) and (\ref{eq:TWAdsScale}), we calculated the excess adsorption per unit area (\ref{eq:Ads}) at various chemical potentials and for four different sized systems, $L/\sigma=500$, $250$, $100$ and $50$. The thickness of the gas film $l_{eq}$ is related to the excess adsorption $\Gamma$ by Eq.\ (\ref{eq:AdsLiq}) or Eq.\ (\ref{eq:AdsGas}) depending on whether the chemical potential of the reservoir is on the liquid or the gas side of bulk coexistence. The coefficients $b_1$ and $b_2$ appearing in the excess grand potential for the system (\ref{eq:GP1}) and adsorption scaling function (\ref{eq:TWAdsScale}) are independent of the precise definition for $l_{eq}$ \cite{NapDiet} but the next to leading order terms with coefficients $c_1$ and $c_2$ are affected by this choice. Figure \ref{fig:adslnliq} displays the DFT results and the effective potential prediction, Eqs.\ (\ref{eq:TWleq}) and (\ref{eq:TWAdsScale}), for the adsorption scaling function on the liquid side of bulk coexistence. There is good agreement between the DFT results and the scaling prediction over a very large range of values of $(L/\sigma)^3\beta\delta\mu$ and we found equally good agreement on the gas side of bulk coexistence (not shown). Near to bulk coexistence, i.e., for small values of $|\delta\mu|$, the liquid-gas interface is close to the mid-point between the two walls so that $|\Gamma|/[A(\rho_l-\rho_g)L]\approx 0.5$. The interface begins to move away from the centre towards one of the walls as $|\delta\mu|$ is increased and the term $\delta\mu(\rho_l-\rho_g)l$ in the grand potential (\ref{eq:GP1}) becomes significant at around $(L/\sigma)^3\beta\delta\mu=10$. Eventually, for large chemical potential difference $|\delta\mu|$, the liquid-gas interface is close to one of the walls and the terms in either $l^{-3}$ or $(L-l)^{-3}$ in the grand potential (\ref{eq:GP1}) become important and the approximation which led to our scaling prediction (\ref{eq:TWleq} and \ref{eq:TWAdsScale}) breaks down. This can be gleaned from our DFT results for $L/\sigma=50$ and $L/\sigma=100$ which deviate slightly from the effective potential prediction at some of the larger values of $(L/\sigma)^3\beta|\delta\mu|$.

\begin{figure}
\centering
\epsfig{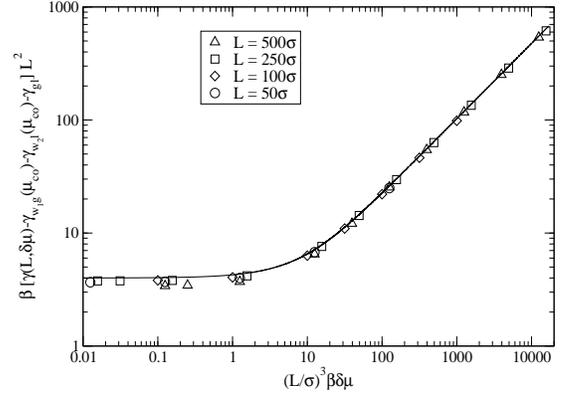}
\caption[Plot to illustrate surface tension scaling in the asymmetric slit on the liquid side of bulk coexistence $\delta\mu>0$]{Plot to illustrate surface tension scaling in the asymmetric slit on the liquid side of bulk coexistence $\delta\mu>0$. Symbols denote DFT results for various wall separations. The independently calculated wall$_1$-gas and wall$_2$-liquid surface tensions at coexistence, and the gas-liquid surface tension have been subtracted from the total surface tension of the system. When multiplied by $L^2$ and plotted against $\delta\mu L^3$ the data for various different $L$ and $\delta\mu$ collapse onto a single curve. The full line is the scaling function $\beta\Sigma\left[\beta\delta\mu(L/\sigma)^3\right]$, valid for $L\rightarrow\infty$, calculated from Eqs.\ (\ref{eq:STscale}) and (\ref{eq:STscalefliq}).}
\label{fig:surliq}
\end{figure}

In Section \ref{sec:ST} we showed that the surface tension of the system can be expressed as the sum of the surface tensions of the three individual interfaces (wall$_1$-gas, gas-liquid and wall$_2$-liquid) plus a term proportional to $L^{-2}$ multiplied by a scaling function $\Sigma [\beta\delta\mu (L/\sigma)^3]$, see Eq.\ (\ref{eq:STscale}). In Figure \ref{fig:surliq} we compare the predicted scaling function (\ref{eq:STscalefliq}) with DFT results obtained on the liquid side of bulk coexistence. For this case and on the gas side of bulk coexistence (not shown) there is good agreement. The very small discrepancy between the DFT results and the prediction from Eq.\ (\ref{eq:STscalefliq}) at small values of $(L/\sigma)^3\beta|\delta\mu|$ probably arises because the difference between the surface tensions, $\gamma(L,\mu)-\gamma_{w_1g}(\mu_{co}) -\gamma_{w_2l}(\mu_{co})-\gamma_{lg}$, is very small compared to the values of the surface tensions themselves. When $L$ is large any error is magnified when multiplied by $L^2$, which explains why the greatest deviation from the theory is for the DFT results at $L=500\sigma$.

\begin{figure}
\centering
\epsfig{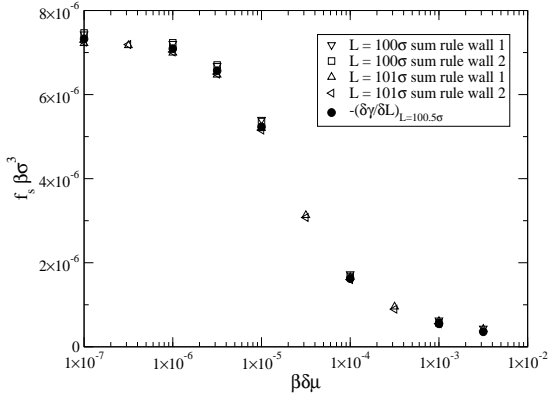}
\caption[DFT results for the solvation force in the asymmetric slit as a function of the deviation from coexistence]{DFT results for the solvation force in the asymmetric slit as a function of the deviation from coexistence $\delta\mu$, for $\delta\mu>0$. The circles ($\bullet$) are data obtained by directly measuring the change in the surface tension between $L=100\sigma$ and $L=101\sigma$. The other symbols are results acquired using the contact density sum rules at the two walls for $L=100\sigma$ and $L=101\sigma$---see Eqs.\ (\ref{eq:Fs1}) and (\ref{eq:Fs2}).}
\label{fig:fs100}
\end{figure}

\begin{figure}
\centering
\epsfig{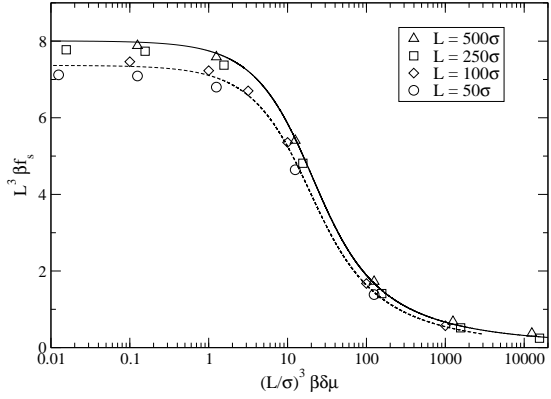}
\caption[DFT results for the solvation force in the asymmetric slit multiplied by $L^3$, obtained using the sum rule at the wet wall]{DFT results for the solvation force in the asymmetric slit multiplied by $L^3$, obtained using the sum rule (\ref{eq:Fs2}) at the wet wall (wall$_2$) for $\delta\mu>0$. The full line is the predicted scaling function $\beta F_s \left[\beta\delta\mu(L/\sigma)^3\right]$, valid as $L\rightarrow\infty$ (\ref{eq:FsScale}). The dashed line is the prediction for $L=100\sigma$ which includes sharp-kink terms of order $L^{-3}$ in the effective potential.}
\label{fig:fsliq}
\end{figure}

There are three different ways that we can extract the solvation force $f_s(L)$ from our numerical data. It can be calculated, using Eq.\ (\ref{eq:f_s}), from the change in surface tension $\delta \gamma$ when the distance between the walls is increased by a small increment $\delta L$, i.e., $f_s(L)=-(\delta\gamma/ \delta L)_{\mu,T,A}$. Alternatively it can be found from the density profile by applying a sum rule at either of the walls, as described in Section \ref{sec:sr}. In Figure \ref{fig:fs100} we plot the solvation force, obtained via these three methods using wall separations of $L=100\sigma$ and $L=101\sigma$, as a function of the chemical potential deviation from bulk coexistence. There is very good agreement between the values for the solvation force obtained using sum rules at the two different walls for each value of $L$. The values calculated from the difference in the surface tension between the systems with $L=100\sigma$ and those with $L=101\sigma$ lie in between the sum rule results from the two separate systems. Having confirmed that the three numerical methods for obtaining $f_s$ are consistent we chose the most convenient to test the scaling behaviour predicted by Eqs.\ (\ref{eq:f_sscale}) and (\ref{eq:FsScale}). For systems on the liquid side of coexistence the solvation force was obtained most easily using the sum rule at the wet wall (\ref{eq:Fs2}). DFT results for $f_s$ at $L=500\sigma$ are close to the predicted scaling function but there is more discrepancy as $L$ is decreased and for $L=50\sigma$ there is a significant deviation---see Fig.\ \ref{fig:fsliq}. This can be explained by the presence of higher order terms (proportional to $L^{-3}$) in the excess grand potential which we neglected in our derivation of Eqs.\ (\ref{eq:f_sscale}) and (\ref{eq:FsScale}). If these terms are included then it is no longer possible to express the solvation force in terms of a scaling function in the manner of Eq.\ (\ref{eq:f_sscale}). In Fig.\ \ref{fig:fsliq} we have plotted the effective potential prediction for the solvation force for $L=100\sigma$ including the next to leading order contributions with their sharp-kink coefficients. The agreement with the DFT results for $L=100\sigma$ is surprisingly good as we expected the coefficients $c_1$, $c_2$ and $c_3$, Eqs.\ (\ref{eq:c1}), (\ref{eq:c2}) and (\ref{eq:c3}) to be affected by the details of the density profile.

\begin{figure}
\centering
\epsfig{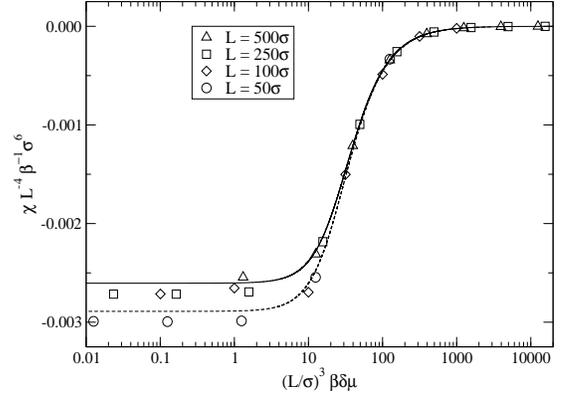}
\caption[Plot to illustrate susceptibility scaling in the asymmetric slit for $\delta\mu>0$] {Plot to illustrate susceptibility scaling in the asymmetric slit for $\delta\mu>0$. The symbols give DFT results for various $L$ and $\delta\mu$, obtained by finding the difference in the adsorption $\Delta \Gamma$ between results at chemical potentials differing by $\beta \Delta \mu=1\times10^{-9}$. The full line gives the predicted scaling function $\beta^{-1} \sigma^6C \left[\beta\delta\mu(L/\sigma)^3\right]$from Eq.\ (\ref{eq:SusScale}), valid as $L\rightarrow\infty$ and the dashed line is a prediction which includes next order terms ($L^{-3}$) in the effective potential approach for $L=100\sigma$.}
\label{fig:SusScaleLiq}
\end{figure}

The effective potential approach predicts a susceptibility which diverges as $L^4$ at bulk liquid-gas coexistence (\ref{eq:SusCo}). Results for the total susceptibility, $\chi$, are shown in Figure \ref{fig:SusScaleLiq} for $\delta\mu>0$. The magnitude of the susceptibility is greatest at bulk coexistence $\delta\mu=0$ when the liquid-gas interface lies at the mid-point between the two walls; in this position the external potential exerted on the fluid by the two walls is very weak and the free energy cost incurred by shifting the interface is smallest. As $|\delta\mu|$ increases the magnitude of $\chi$ begins to fall sharply as the liquid-gas interface moves away from the centre of the slit---at around $(L/\sigma)^3\beta\delta\mu=10$. The DFT results are in agreement with the predicted scaling function (\ref{eq:SusScale}) for $L=500\sigma$ but begin to deviate for smaller $L$. Including higher order terms (those proportional to $l^{-3}$, $(L-l)^{-3}$ and $L^{-3}$ in the effective potential (\ref{eq:GP1}), with sharp-kink coefficients) does account for some of the disparity, as is illustrated by the prediction for $L=100\sigma$ (dashed line in Fig.\ \ref{fig:SusScaleLiq}) which is much closer to the corresponding DFT results. The remaining discrepancy is probably due to the coefficients of these terms being modified because the DFT density profiles are smooth rather than sharp--kink--like.

\begin{figure}
\centering
\epsfig{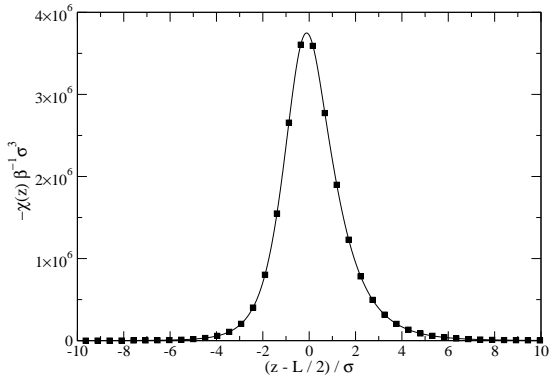}
\caption[The magnitude of the local susceptibility near the midpoint between the two walls]{The magnitude of the local susceptibility $-\chi(z)\equiv -\chi(z;L,\mu)=\left(\partial \rho(z)/\partial \mu\right)_{L,T}$ near the midpoint between the two walls obtained by subtracting profiles at $\beta\mu=\beta\mu_{co}$ and  $\beta\mu=\beta\mu_{co}+1\times10^{-10}$ for a wall separation of $L=250\sigma$ ($\blacksquare$). The line gives the product $-(\partial l_{eq}/\partial \mu)_{L,T}(\partial \rho(z)/\partial z)$.}
\label{fig:locsus}
\end{figure}

Figure \ref{fig:locsus} shows the local susceptibility $\chi(z;L,\mu)$ defined by Eq.\ (\ref{eq:LocSusDef}), i.e., the change in the density profile $\rho(z)$ with chemical potential, for a wall separation of $250\sigma$, evaluated at bulk liquid-gas coexistence $\mu_{co}$. The peak in the magnitude of the susceptibility coincides with the position of the liquid-gas interface lying at the centre of the slit confirming that the most significant change in the density profile as the chemical potential is varied involves the displacement of this interface. This conclusion is backed up by the similarity between the local susceptibility $-(\partial \rho(z)/\partial \mu)_{L,T}$ and the product $(\partial l_{eq}/\partial \mu)_{L,T}(\partial \rho(z)/\partial z)$, see Eq.\ (\ref{eq:LocSus}), which is also plotted in Fig.\ \ref{fig:locsus}. (The quantity $\left(\partial l_{eq}/\partial \mu\right)_{L,T}$ was calculated from the change in the adsorption using $\left(\partial l_{eq}/\partial \mu\right)_{L,T}=-(\partial \Gamma/\partial \mu)_{L,T}[A(\rho_l-\rho_g)]^{-1}$.)

\begin{figure}
\centering
\epsfig{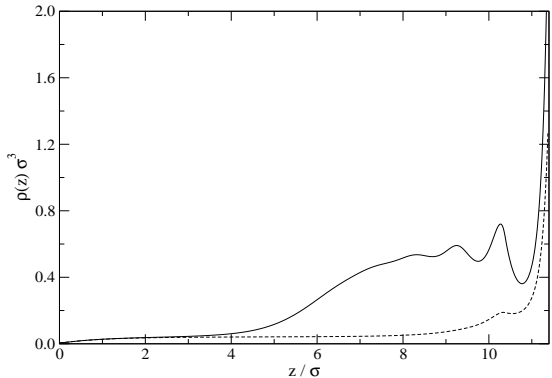}
\caption[The coexisting density profiles of the `delocalised interface state' and the evaporated state for $L^{co}_{evap}=11.4\sigma$ at $\delta\mu=0$]{The coexisting density profiles of the `delocalised interface state' (full line) and the evaporated state (dashed line) for $L^{co}_{evap}=11.4\sigma$ and $\delta\mu=0$.}
\label{fig:profevap11_4}
\end{figure}

In Section \ref{sec:cap} we predicted that a first order capillary evaporation transition could occur in our asymmetric slit and used the effective potential approach to calculate the chemical potential at this transition as a function of wall separation (Fig.\ \ref{fig:EvapOpp}). The line of evaporation transitions crosses bulk liquid-gas coexistence at a predicted value of $L_{evap}^{co}=13\sigma$, obtained from Eq.\ (\ref{eq:Levap}) using sharp-kink values for the coefficients $b_1$, $b_2$, $b_3$ and $b_g$ and DFT results for the surface tensions $\gamma_{w_2g}^*(\mu_{co})$, $\gamma_{w_2l}(\mu_{co})$ and $\gamma_{lg}$. (Note once again $\gamma_{w_2g}^*(\mu_{co})$, is the excess grand potential of the wall$_2$-thin film state which is metastable at $\mu=\mu_{co}$.) Comparing the excess grand potential (calculated from DFT density profiles in the asymmetric slit) for the evaporated state with that of the state with that of the `delocalized interface state', as a function of $L$ gave the slightly smaller value of $L_{evap}^{co}=11.4\sigma$. It is not surprising that the effective potential prediction of $L_{evap}^{co}$ is not particularly accurate at such small wall separations. Figure \ref{fig:profevap11_4} shows the density profiles of the two coexisting states at $L_{evap}^{co}$ and it is clear that the density profile for the delocalised interface state is far from the sharp-kink approximation for the profile which forms the basis of the effective potential treatment in Section \ref{sec:EIP}.

\begin{figure}
\centering
\epsfig{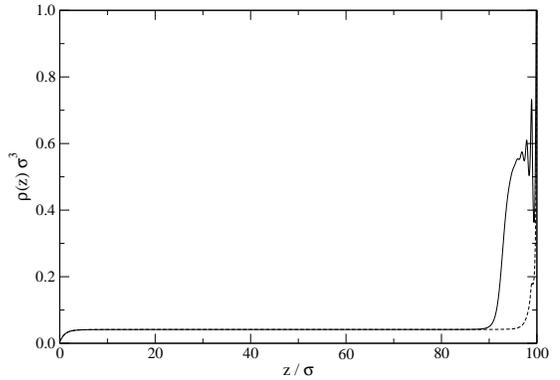}
\caption[The coexisting density profiles of the `delocalised interface state' and the evaporated state for $L=100\sigma$]{The coexisting density profiles of the `delocalised interface state' (full line) and the evaporated state (dashed line) for $L=100\sigma$ and $\beta\delta\mu_{evap}=-4.1344\times10^{-3}$. The profiles are almost identical to those corresponding to the coexisting thick and thin films at the prewetting transition for a single solvophilic wall$_2$.}
\label{fig:profevap100}
\end{figure}

At larger wall separations, $L>L_{evap}^{co}$, capillary evaporation occurs when the system is on the gas side of bulk coexistence, i.e., $\delta\mu_{evap}<0$ (see Fig.\ \ref{fig:EvapOpp}). This is in contrast to the situation for a system with identical drying walls where capillary evaporation may occur only as bulk coexistence is approached from the {\em {liquid}} side (see Fig.\ \ref{fig:EvapId}). For wall separations $L\gtrsim50$ evaporation is predicted to occur at chemical potentials very close to $\delta\mu_{pw}$, the chemical potential at the prewetting transition at the isolated solvophilic wall. Using DFT we found that $\beta\delta\mu_{pw}=-4.1368\times10^{-3}$ for prewetting at (single) wall$_2$, c.f., the value $-3.4\times10^{-3}$ from the effective potential approach used in Fig.\ \ref{fig:EvapId}. The chemical potential at the evaporation transition for a wall separation of $L=100\sigma$ was found to be $\beta\delta\mu_{evap}=-4.1344\times10^{-3}$ which is indeed very close to $\delta\mu_{pw}$. Figure \ref{fig:profevap100} displays the coexisting density profiles at capillary evaporation for $L=100\sigma$. Unlike the profiles shown in Fig.\ \ref{fig:profs100dmu} for smaller values of $\lvert\delta\mu\rvert$, at this chemical potential the gas-liquid interface in the `delocalised interface state' has moved away from the centre of the slit and closer to the solvophilic wall$_{2}$. The evaporated state is filled with gas apart from a thin film (thickness $\approx 1 \sigma$) of higher density fluid next to the solvophilic wall. The two density profiles in Fig.\ \ref{fig:profevap100} are, apart from the region very close to the solvophobic wall, almost identical to those of the coexisting thick and thin films at the prewetting transition at wall$_2$.

\section{Discussion}
\label{sec:distw}
In this paper we made a detailed investigation of the `delocalised interface phase' or `soft mode phase', identified originally by Parry and Evans \cite{ParryEv} for a model fluid described by long-range $(-r^{-6})$ dispersion potentials which is confined between competing planar substrates. 
We summarize our main results as follows:
\begin{enumerate}
\item
Using an effective potential approximation (Section \ref{sec:EIP}) we derived scaling forms for the Gibbs adsorption $\Gamma(L,\mu)$, the surface tension $\gamma(L,\mu)$, the total susceptibility $\chi(L,\mu)$ and the solvation force $f_s$ that depend on the dimensionless product $(L/\sigma)^3\beta\delta\mu$, where the power of three reflects directly the power of $-6$ in the interparticle potentials. The scaling forms were confirmed by numerical results from calculations based on a non-local DFT (Figs.\ \ref{fig:adslnliq}, \ref{fig:surliq}, \ref{fig:fsliq} and \ref{fig:SusScaleLiq}). Provided the wall separation $L$ is sufficiently large, typically $L \gtrsim 100 \sigma$, there is good agreement with the explicit scaling predictions. Perhaps most striking is the prediction that $\chi(L,\mu_{co}) \sim L^4$---see Eq.\ (\ref{eq:SusCoAs}). This result corresponds to a transverse correlation length that diverges as $\xi_{||}\sim L^2$, for $L\rightarrow\infty$ \cite{ParryEv}. Both these results imply that the central liquid-gas interface undergoes extremely large interfacial fluctuations, correlated over very long distances parallel to the walls. These interfacial fluctuations reflect the smallness of the free--energy cost of shifting the whole liquid-gas interface and are characteristic of the `soft mode phase'. Recall that for $\delta\mu=0$ the liquid-gas interface is constrained by only the tails of the interfacial potentials which decay as a power-law ($l^{-2}$ for dispersion forces). By contrast, for short range forces the decay is exponential which leads to the result quoted in Section \ref{sec:int}: $\chi (L,\mu_{co}) \sim e^{ \kappa L/2}$, for $L\rightarrow\infty$ \cite{ParryEv}.

\item
The central liquid-gas interface is subject to an effective repulsive force from each wall as wall$_1$ endeavours to become dry and wall$_2$ to wet. The resulting solvation force is repulsive. For dispersion forces we find $f_s\sim +L^{-3}$ for $\delta\mu=0$. Specifically, in the perfectly antisymmetric case with $b_1=b_2$, the liquid-gas interface lies in the centre of the slit and from Eq.\ (\ref{eq:FsScale})
\begin{equation}
f_s(L,\mu_{co})=\frac{1}{L^3}\left(16b_2+2b_3\right).
\label{eq:fsco}
\end{equation}
This result should be contrasted with that for short-range forces in the `soft mode phase' where $f_s\sim+e^{-\kappa L/2}$, for $L\rightarrow\infty$ \cite{ParryEv}. Employing the sum rules of Section \ref{sec:sr} we calculated the solvation forces from DFT using three different routes, demonstrating their consistency and the accuracy of our numerics---see Fig.\ \ref{fig:fs100}.

It is interesting to compare our results for $f_s$ with those obtained using an equivalent DFT approach for a different model fluid confined between two identical walls \cite{MacDrzBry}. In this study the wall--fluid potentials decayed as $z^{-3}$ and the fluid--fluid pair potentials were short-ranged, i.e., truncated Lennard--Jones. Away from any phase transitions and bulk criticality the solvation force was found to decay as $f_s\sim +L^{-3}$. However, including the full $-r^{-6}$ tails of the fluid--fluid pair potential results in an additional {\em attractive} $-L^{-3}$ contribution to $f_s$ \cite{MacDrzBry}. The asymptotic decay of the solvation force for a system with identical walls can be either attractive or repulsive depending on the choice of parameters in the (power-law) wall--fluid and fluid--fluid potentials and on the thermodynamic state point.

\item
In Section \ref{sec:cap} we used the effective interfacial potential approximation to investigate possible phase behaviour for our model fluid confined between asymmetric walls. Figure \ref{fig:EvapOpp} shows the predicted phase diagram for the single temperature $T=0.8T_c$ that we studied using DFT in Section \ref{sec:DFTres}. Depending on the wall separation $L$ capillary evaporation, i.e., a transition to a dilute gas state with only a thin adsorbed film of liquid--like density at wall$_2$, can occur not only for $\delta\mu>0$ (the situation that pertains for capillary evaporation in the case of identical walls) but also for $\delta\mu<0$. Fig.\ \ref{fig:profevap11_4} shows the coexisting density profiles calculated from DFT when capillary evaporation occurs at $\delta\mu=0$, i.e., for the wall separation $L^{co}_{evap}=11.4\sigma$. Evaporation occurs because for smaller wall separations, $L<L_{evap}^{co}$, the free energy cost from the interface interactions plus the surface tension of the central liquid-gas interface is so great that it is more favourable for the slit to be filled by the evaporated `gas', despite the fact that the surface tension $\gamma_{w_2g}^*(\mu)$ between gas and (solvophilic) wall$_2$ is large. For $L>L_{evap}^{co}$ capillary evaporation occurs on the gas side of bulk coexistence, i.e., $\delta\mu<0$. However, in this regime for large $L$ the evaporation we observe is closely connected to the prewetting transition that occurs at an isolated $(L=\infty)$ solvophilic wall. Thus for $L=100\sigma$ DFT yields $\beta(\delta\mu_{evap}-\delta\mu_{pw})=2\times10^{-6}$ and Fig.\ \ref{fig:profevap100} displays the coexisting density profiles at this evaporation transition. These profiles are almost identical to those at the prewetting transition. Our present study describes evaporation, rather than condensation, because the solvophilic wall$_2$ undergoes a first order wetting transition at $T_W<0.8T_C$ whereas the purely repulsive solvophobic wall$_1$ is dry at all temperatures between the bulk triple and critical temperatures; there is no pre-drying transition. In a perfectly antisymmetric system, such as an Ising model subject to equal but opposite surface fields ($h_{D}=-h_1$) for which wall$_1$ undergoes a first order `drying' transition at the same temperature as wall$_D$ undergoes the equivalent wetting transition, the evaporated state with a thin film of `liquid' at wall$_D$ would coexist with the condensed state with a thin film of gas at wall$_1$ and with the `delocalised interface state' at bulk coexistence $\delta\mu=0$, resulting in a triple point at $L=L_{cond}^{co}=L_{evap}^{co}$. The admirable review by Binder et.al. \cite{BinLanMul} provides an illuminating summary of the genesis of such triple points in Ising models, focusing mainly on temperature dependence at fixed $L$ rather than fixing $T$ and varying $L$ as we do here.

M\"uller et.al (see Ref.\ \cite{MulBin} for a summary) also found triple points in their self-consistent field treatments of symmetric AB binary polymer blends confined between perfectly antisymmetric walls, i.e., one wall attracts the A component with exactly the same strength as the other wall attracts component B and there is a first-order wetting transition. Although the interparticle potentials in these studies are short-ranged it is probable than many of the gross features of the phase diagram found by M\"uller et.al. will also apply to our system.
\end{enumerate}
Our results pertain to non-retarded (London) dispersion forces: we deliberately chose to consider interparticle potentials that decay as $-r^{-6}$ since these describe the models that are conventionally used in the physics of liquids. In the physics of wetting it is well-known that phase transitions at a single wall ($L=\infty$) depend profoundly on whether the interparticle potentials are short--ranged or long--ranged (power--law) \cite{DietR}. Indeed the critical exponents describing continuous wetting transitions depend on the specific power. Relevant to the present study is the observation that for fluids subject to wall-fluid potentials decaying as $-z^{-(n+1)}$, corresponding to a wall particle-fluid particle potential decaying as $-r^{-(n+4)}$, the leading term in the excess grand potential (\ref{eq:GP}) goes as $l^{-n}$, where $n=2$ for non-retarded dispersion forces. For the complete wetting transition (from off bulk coexistence, with $\delta\mu\rightarrow0$) the upper critical dimension for systems with power-law potentials is given by \cite{DietR,ParryEv}:
\begin{equation}
d^*=3-4(n+2)^{-1}
\label{eq:ucd}
\end{equation}
Thus for complete wetting in the presence of non-retarded dispersion forces $d^*=2$ and mean-field predictions for the critical exponents will remain valid for all dimensions $d>2$. We recall now the heuristic scaling ansatz in Section 4 of Parry and Evans \cite{ParryEv}. This argues that the only relevant scaling argument in the `soft mode phase' must be the ratio $L/(2l_{eq})$, where the equilibrium thickness of the (drying) film is given by the complete wetting result $l_{eq} \sim \lvert \delta\mu\rvert ^{-\beta_s^{co}}$ with $\beta_s^{co}=(n+1)^{-1}$ for power--law potentials. The idea is that the maximum thickness of the film at equilibrium is $L/2$, at $\delta\mu=0$, and that the film thins according to the complete wetting behaviour. Note that there is no temperature scaling variable: it is assumed that the temperature is sufficiently far below $T_C$ and sufficiently far above the interface delocalization temperature that it is irrelevant. Only the chemical potential deviation $\delta\mu$ is relevant for scaling. The resulting scaling predictions \cite{ParryEv} for a general power-law potential are $\chi(L,\mu_{co})\sim L^{n+2}$ and $\xi_{||}\sim L^{(n+2)/2}$, for $\delta\mu=0$ and $L\rightarrow\infty$.The solvation force is predicted to decay as $+L^{-(n+1)}$ in this `soft-mode' regime.

Our present results, based on the effective interfacial potential treatment of Section \ref{sec:EIP} and the DFT calculations of Section \ref{sec:DFTres}, are evidently consistent with the scaling ansatz for the particular case $n=2$. Clearly both are mean-field treatments; they both fail to capture all of the effects of capillary-wave fluctuations. Nevertheless, if the heuristic scaling ansatz of Ref.\ \cite{ParryEv} is correct, and there seems no reason to doubt its validity, and given Eq.\ (\ref{eq:ucd}) for the upper critical dimension for complete wetting, it follows that for all power--law potentials ($n$ finite) $d^*<3$ so that the mean-field predictions should be valid as regards the dependence of thermodynamic quantities and the correlation length on the wall separation $L$, for $L\rightarrow\infty$. This observation is important since it implies that the leading order scaling functions that we derived from the (mean-field) effective interfacial potential with $n=2$, and checked using the microscopic DFT approach, should remain valid in the presence of fluctuations. Moreover, so should the corresponding analysis for any finite $n$. Of course, in an exact treatment the coexisting bulk densities $\rho_l$ and $\rho_g$ entering the Hamaker constants $b_1$, $b_2$ and $b_3$---see Eqs.\ (\ref{eq:b1}, \ref{eq:b2} and \ref{eq:b3})---should be replaced by their exact counterparts for the model fluid under consideration.

What is omitted in our mean--field DFT approach? Capillary--wave fluctuations must broaden the central liquid--gas interfacial density profile in the `soft mode phase'. This effect is not captured within the simple DFT employed here, i.e., the results shown in Fig.\ \ref{fig:profs100dmu} showing rather sharp interfaces do not encompass such broadening. Computer simulations for the same model would exhibit broader interfacial density profiles. In spatial dimension $d=3$ the extent of the capillary wave broadening is given by $\xi_{\perp}\sim (\ln \xi_{||})^{1/2}$. For the large wall separations that we consider this broadening is not insubstantial and is certainly on the scale of the intrinsic width that emerges from our DFT calculations. Once again we contrast this result with the case of short-ranged forces where $\xi_{||}\sim e^{\kappa L/4}$ \cite{ParryEv} and thus $\xi_{\perp}\sim L^{1/2}$, for $\delta\mu=0$ and $L\rightarrow \infty$. This last result was confirmed (for relatively small $L$) by measurements of the width of the magnetization profile in Ising model simulations, e.g., the summary in Section 3.4 of Ref.\ \cite{BinLanMul}.

We emphasize that the analysis in the present paper is for a single temperature that lies in the `delocalised interface phase'. Unlike earlier studies that focused on systems with short-ranged forces \cite{ParryEvPRL,ParryEv,BinLanFerPRL,BinLanFerPRE,BinEvLanFer,BinLanMul,MulBin} we have not attempted to investigate the localisation-delocalisation transition that is intimately linked to the wetting transition at an (isolated) confining wall. In the original studies \cite{ParryEvPRL,ParryEv} the wetting transition was continuous (critical) and there were detailed predictions for the location of the critical point and the nature of the criticality in the confined system with perfectly antisymmetric walls. Subsequently these were confirmed by results from Ising simulations \cite{BinLanFerPRL,BinLanFerPRE,BinEvLanFer,BinLanMul}. For perfectly antisymmetric walls the critical temperature $T_{CL}$ (of the interface delocalization--localisation transition) lies below the single wall critical wetting temperature $T_W$ for any finite $L$ and in the limit $L\rightarrow \infty$, $T_{CL}\rightarrow T_W$. In the present analysis we deliberately chose the wall--fluid interaction parameters in order that the walls were antisymmetric to leading order in the excess grand potential (\ref{eq:GP}). Specifically we set $b_1$, the Hamaker constant for drying at wall$_1$ to be equal to $b_2$, the Hamaker constant for wetting at wall$_2$. Of course, the Hamaker constants depend on temperature through their dependence on the bulk coexisting densities, see Eqs.\ (\ref{eq:b1}) and (\ref{eq:b2}). $b_1$ and $b_2$ are equal at only one temperature. At a lower temperature these quantities are not identical and the system is not antisymmetric. This makes locating the localization--delocalisation transition rather demanding.

We have also deliberately avoided the bulk critical regime, near $T_c$, where the bulk correlation length $\xi_b\sim L$. For competing walls we expect a universal `critical Casimir' solvation force that decays as $f_s\sim +L^{-d}$ exactly at criticality. Thus for d=3 the power--law decay of the critical Casimir force is the same as that found for non-retarded dispersion forces in the `soft mode phase' below $T_c$. The amplitudes are, of course, different. Ascertaining the precise nature of cross--over as $T$ approaches $T_c$ from below is challenging.

We should comment on studies of other model fluids in asymmetrical confinement. As we mentioned in Section \ref{sec:int}, colloid--polymer mixtures are a particularly attractive system to study both from an experimental and a theory/simulation perspective---especially if one adopts the simplest Asakura-Oosawa-Vrij (AO) model for these binary mixtures \cite{BinHorVinVir}. Within the context of this fluid model, a hard--wall will favour wetting by the colloid--rich phase, as a result of depletion interactions, while a suitably tailored wall, with appropriate coating, can be made to favour the polymer rich phase \cite{BinHorVinVir}. Since depletion interactions arise from excluded volume considerations they are short--ranged. Indeed they are strictly finite ranged in the case of the AO model. (Experimentally it is assumed there are no residual dispersion forces, i.e., refractive index matching is perfect.) Nevertheless, several of the results obtained by De Virgiliis et.al. \cite{VirVinHorBin,VirVinHorBinPRE} in simulations of the AO model are close to those that we find for our atomic model. For example, the density profiles for the colloid species in the `delocalised interface phase' (see Figs.\ 2b and 15d, where $L=10$ colloid diameters,  of Ref.\ \cite{VirVinHorBinPRE}) are very similar to what we find in the corresponding `delocalised interface state' (see Fig.\ \ref{fig:profevap11_4} of the present). It is expected that several features the phase behaviour ascertained for the AO model in asymmetric confinement will pertain to the present fluid model. Of course, the details of the decay (or divergence) of thermodynamic functions will be different.

Returning to the present one-component model one might argue that the choice of a purely repulsive substrate (wall$_1$) is inappropriate for any real fluid. In reality there is always some residual attractive dispersion interaction between the fluid and the substrate and Young's contact angle is always $<\pi$; one does not have complete drying. However, one can easily circumvent this objection by considering a binary mixture with dispersion forces. Similar to the case of polymer blends, one can envisage a binary molecular mixture confined by asymmetric walls that favour one species or the other so that in a certain temperature range one wall can be completely wet by the phase rich in species A and the other completely wet by the phase rich in species B. It is straightforward to extend the present effective interfacial potential approach to the binary case and evaluate the appropriate Hamaker constants that determine the scaling functions. The phenomenology is somewhat richer because we deal with a mixture but the basic scaling predictions in the `soft mode phase' remain the same.

It is important to review the length scales that we considered in this study. In order to confirm the scaling predictions it was necessary to perform DFT calculations for extremely large wall separations---in some cases up to $250\sigma$. Clearly this range is far beyond the current realms of molecular simulation. Since the 1930s it has been recognized that dispersion forces play a key role in the physics of wetting and in many other aspects of confinement. It is clear that a full understanding of several subtle aspects of wetting phenomena requires a microscopic approach that incorporates these long--ranged forces and that treats short--ranged correlations, arising from packing of the `particles', in a realistic fashion. Our present study and our earlier one on wetting/drying at a curved substrate \cite{StewEv} demonstrate that classical DFT provides a successful approach, albeit one that treats the attractive interactions at mean-field level. One should also note that retardation becomes relevant in real fluids and for real substrates. For sufficiently large inter--nuclear separation $r$ the atom--atom pair potential crosses over from $-r^{-6}$ to $-r^{-7}$ decay. This implies cross over to ultimate $-z^{-4}$ decay of the wall--fluid potential and to a leading contribution of $l^{-3}$ in the excess grand potential (\ref{eq:GP}). Were we to attempt a more realistic treatment of interactions for large separations, where retarded forces are relevant, the ultimate power laws would be different from those obtained for non-retarded dispersion forces. For example, the solvation force decays ultimately as $f_s\sim +L^{-4}$, rather than as $+L^{-3}$. This observation is important since we showed in our microscopic calculations for the non--retarded case (see Fig.\ \ref{fig:fsliq}) that agreement with the scaling limit is achieved accurately only for very large $L$, say $>100 \sigma$. In real systems retardation will kick in at similar separations. 

Can the results of our present study be related to experiment? By measuring the force between the tip and a substrate in an atomic force microscope or between the crossed cylinders in a surface force apparatus (SFA) one can obtain information about the solvation force. In particular, by using the Derjaquin approximation, one can relate the force measured in the SFA to $f_s(L,\mu)$, the quantity we calculated and which pertains to two (infinite area) parallel walls \cite{Isr}. Our present study was motivated in part by an experimental study \cite{ZhaZhuGran} that used a SFA to investigate both the normal force and the response to shear deformation of water, in contact with a reservoir at normal temperature and pressure, confined between hydrophobic and hydrophilic surfaces. The title of Ref.\ \cite{ZhaZhuGran} refers to a ‘Janus Interface’. In the experiments the hydrophobic surface was either a cylinder of mica coated with a self-assembled monolayer of octadecyltriethoxysilane (OTE) or mica with a thin film of silver then coated with a self-assembled octadecanethiol layer. The hydrophilic surface was untreated mica which the authors assumed to be wet completely by water. It was observed that the shear response was extraordinarily noisy and indicated a distribution of relaxation processes. The authors mention `giant fluctuations (of the dynamical shear responses)' and allude to the work of Parry and Evans \cite{ParryEvPRL}. Without being explicit they appear to infer that their experimental configuration corresponds to that of the `soft-mode phase' in Ref.\ \cite{ParryEvPRL,ParryEv} and they write about a `flickering, fluctuating complex' in which the capillary wave fluctuations are somehow thwarted.

\renewcommand{\labelenumi}{(\roman{enumi})}
Two observations are relevant in the light of our present analysis of the soft--mode phase:
\begin{enumerate}
\item
The measured static force-distance profile (see inset to Fig.\ 1 in \cite{ZhaZhuGran}) is attractive over a wide range of surface separations, from $1000$ down to about $10$ molecular (water) diameters. This contrasts sharply with what we calculate for fluid confinement between competing solvophobic (complete drying) and solvophilic (complete wetting) walls. For this case we find the solvation force is repulsive at large separations throughout the region of the soft-mode phase. At small separations $L$, where the confined fluid is a dense liquid, the solvation force can oscillate as a function of $L$ due to packing effects.
\item
The hydrophobic surfaces prepared in \cite{ZhaZhuGran} are, of course, not completely dry. The contact angle measured for the OTE surface is $110\pm 2^o$ and that for the thiol surface is $120\pm 2^o$. Thus the situation realized in the experiment \cite{ZhaZhuGran} does not match that described in \cite{ParryEvPRL,ParryEv} and which is considered in detail in this paper. It difficult to see why the particular choice of wet and partially dry surfaces studied in the SFA experiments \cite{ZhaZhuGran} could give rise to a wildly fluctuating liquid--gas interface---see also the comments of Pertsin and Grunze \cite{PertGrun}. Nevertheless the experimental observations remain intriguing and worthy of further investigation.
\end{enumerate}
\section{Acknowledgements}
We thank R. Roth for helpful advice on numerical methods in the early stage of this work. MCS is grateful to EPSRC for financial support during her PhD studies. The research described here is based on part of her PhD thesis (University of Bristol 2006).

\appendix
\section{The wall--fluid potential}
We model the substrate as a block of (wall) particles of constant, uniform density $\rho_w$. The wall cannot be penetrated by the fluid particles, i.e., the wall-particle interaction potential is infinitely repulsive at the plane of contact. This hard-core repulsion between the particles results in an excluded volume, of width $dw$, where the density vanishes, at the boundary between the wall and the fluid, as illustrated in Fig.\ \ref{fig:walls}. The wall is structureless, i.e., parallel to the interface the density is taken to be uniform. The attractive potential exerted by the wall on the fluid particles can be calculated by first considering the interaction potential between individual wall particles with fluid particles. This is chosen to be of the same form as the fluid-fluid inter-particle potential. The attractive part is given by
\begin{equation}
\phi_{att}^{wf}(r_{12}) =
\begin{cases}
4\epsilon_{wf}\left[\left(\frac{\sigma_{wf}}{r_{12}}\right)^{12}-\left(\frac{\sigma_{wf}}{r_{12}}\right)^6\right]\quad &r_{12}>r_{min,wf} \\
-\epsilon_{wf} &r_{12}<r_{min,wf}
\end{cases}
\label{eq:Lennard_Jones_Potential_Wall}
\end{equation}
(c.f.\ Eq.\ \ref{eq:Lennard_Jones_Potential}) where $r_{12}$ is the distance between the wall particle and the fluid particle and $\epsilon_{wf}$ and $\sigma_{wf}$ are the strength and range of the wall-fluid potential, respectively and $r_{min,wf}=2^{1/6}\sigma_{wf}$. The total potential, exerted by the entire wall, on a single fluid particle at a perpendicular distance $z$ from the contact plane (see Fig.\ \ref{fig:walls}) is found by integrating the inter-particle potential over the semi-infinite volume of the wall:
\begin{eqnarray}
V_w(z)=
\begin{cases}
\infty, \phantom{\rho_w\int_{z+dw}^{\infty}{{\rm d}z' \int_0^{\infty}{2\pi r\phi_{att}^{wf}(\sqrt{r^2+z'^2}) \, {\rm d}r}}} z<0,
\\
\rho_w\int_{z+dw}^{\infty}{{\rm d}z' \int_0^{\infty}{2\pi r\phi_{att}^{wf}(\sqrt{r^2+z'^2}) \, {\rm d}r}}, \phantom{\infty}  z>0. \nonumber
\end{cases}
\label{eq:wall_potential_int}
\end{eqnarray}
We obtain
\begin{eqnarray}
V_w(z)=
\begin{cases}
\infty, \phantom{\rho_wv_{wf}(z+dw)} \hspace{1mm} z<0,
\\
\rho_wv_{wf}(z+dw), \phantom{\infty} \hspace{1mm} z>0,
\end{cases}
\label{eq:wall_potential_app}
\end{eqnarray}
where the function $v_{wf}(z')$ is found by replacing $\epsilon$, $r_{min}$ and $\sigma$ by $\epsilon_{wf}$, $r_{min,wf}$ and $\sigma_{wf}$ in Eq.\ (\ref{eq:Potentialvz}) below.
\section{The interaction between two interfaces}
\label{sec:intint}
For long-ranged interparticle forces the dominant contribution to the interaction between two interfaces separated by distance $l$ is from the tails of the interparticle potentials, which decay as inverse powers of distance. Below we calculate the interaction potential $\omega(l)$ between the wall$_1$-gas and gas-liquid interfaces, that is the extra free energy per unit area of interface for a layer of gas thickness $l$, compared to the free energy for the two separate interfaces ($l\rightarrow\infty$). An equivalent procedure may be followed to calculate the interactions between the other pairs of interfaces. Using the sharp-kink approximation, in which the fluid densities are taken to be uniform and equal to the bulk densities $\rho_g$ and $\rho_l$, respectively, on each side of the sharp gas-liquid interface, we find
\begin{widetext}
\begin{equation}
\omega(l)=(\rho_l-\rho_g)\left(\rho_{w1}\int_{l+dw}^{\infty}{v_{wf}(z') \,{\rm d}z'}-\rho_g\int_l^{\infty}{v(z') \,{\rm d}z'}\right)
\label{eq:wl_app}
\end{equation}
where $\rho_gv(z')$ is the potential due to a semi-infinite slab of the gas a distance $z'$ away:
\begin{equation}
\rho_gv(z') =  \rho_g\int_{z'}^{\infty}{{\rm d}z \int_0^{\infty}{2\pi r\phi_{att}(\sqrt{r^2+z^2}) \, {\rm d}r}}.
\label{eq:IntPotv}
\end{equation}
Performing the integration we obtain

\begin{equation}
v(z') = 
\begin{cases}
4\pi \epsilon\left(\frac{\sigma^{12}}{45z'^{9}}-\frac{\sigma^{6}}{6z'^{3}}\right) & z'>r_{min}
\\
4\pi\epsilon\left(\frac{r_{min}^2z'}{4}-\frac{z'^3}{12}-\frac{r_{min}^3}{6}+\frac{2\sigma^{12}}{9r_{min}^9}-\frac{2\sigma^6}{3r_{min}^3}-\left[\frac{\sigma^{12}}{5r_{min}^{10}}-\frac{\sigma^6}{2r_{min}^4}\right]z'\right) \quad &z'<r_{min}.
\end{cases}
\label{eq:Potentialvz}
\end{equation}
Similarly, $\rho_{w1}v_{wf}(z')$ is the potential due to a semi-infinite slab of the wall a distance $z'$ away where $v_{wf}(z')$ is found by replacing $\epsilon$, $r_{min}$ and $\sigma$ by $\epsilon_{wf}$, $r_{min,wf}$ and $\sigma_{wf}$ in Eq.\ (\ref{eq:Potentialvz}). Integration of (\ref{eq:wl_app}) gives
\begin{eqnarray}
\omega(l)&=&\left(\rho_g\epsilon\sigma^6-\rho_{w1}\epsilon_{wf}\sigma_{wf}^6\right)\frac{b_0}{l^2}+2dw\rho_{w1}\epsilon_{wf}\sigma_{wf}^6\frac{b_0}{l^3}+O\left(\frac{dw^2}{l^4}\right)
\\
&=&\frac{b(T)}{l^2}+\frac{c(T)}{l^3}+O\left(\frac{1}{l^4}\right),
\label{eq:w_app}
\end{eqnarray}
\end{widetext}
where the temperature dependent coefficients $b(T)$ and $c(T)$ are given by
\begin{equation}
b(T)=\left(\rho_g\epsilon\sigma^6-\rho_{w1}\epsilon_{wf}\sigma_{wf}^6\right)b_0
\label{eq:b_app}
\end{equation}
and
\begin{equation}
c(T)=2dw\rho_{w1}\epsilon_{wf}\sigma_{wf}^6b_0
\label{eq:c_app}
\end{equation}
where $b_0=(\rho_l-\rho_g)\frac{\pi}{3}$.

\bibliography{thesis_bibfile}

\end{document}